\documentclass[useAMS,usenatbib]{mn2e}

\usepackage{times}
\usepackage{graphicx}
\usepackage{subfigure}

\newcommand{\gsim}{\mathrel{\hbox{\rlap{\lower.55ex \hbox {$\sim$}}
                   \kern-.3em \raise.4ex \hbox{$>$}}}}
\newcommand{\lsim}{\mathrel{\hbox{\rlap{\lower.55ex \hbox {$\sim$}}
                   \kern-.3em \raise.4ex \hbox{$<$}}}}
\newcommand{\msun}{\mbox{M$_\odot$}}

\title[Star cluster evolution following gas dispersal]{On the evolution of a star cluster and its multiple stellar systems following gas dispersal}

\author[N. Moeckel \& M. R. Bate]
  {Nickolas Moeckel$^{1,2,3}$\thanks{E-mail: moeckel@ast.cam.ac.uk} and Matthew R. Bate$^2$\thanks{E-mail: mbate@astro.ex.ac.uk}\\
  $^1$SUPA, School of Physics and Astronomy, University of St Andrews, North Haugh, St Andrews, Fife, KY16 9SS \\
  $^2$School of Physics, University of Exeter, Stocker Road, Exeter EX4 4QL \\
  $^3$Institute of Astronomy, University of Cambridge, Madingley Road, Cambridge CB3 0HA 
}

\date{\today}

\begin{document}
\maketitle

\begin{abstract}
We investigate the evolution, following gas dispersal, of a star cluster produced from a hydrodynamical calculation of the collapse and fragmentation of a turbulent molecular cloud.  We find that when the gas, initially comprising $\approx 60$\% of the mass, is removed, the system settles into a bound cluster containing $\approx 30-40$\% of the stellar mass surrounding by an expanding halo of ejected stars.  The bound cluster expands from an initial radius of $<0.05$~pc to $1-2$~pc over $\approx 4-10$~Myr, depending on how quickly the gas is removed, implying that stellar clusters may begin with far higher stellar densities than usually assumed.  With rapid gas dispersal the most massive stars are found to be mass segregated for the first $\sim 1$~Myr of evolution, but classical mass segregation only develops for  cases with long gas removal timescales.  Eventually, many of the most massive stars are expelled from the bound cluster.  Despite the high initial stellar density and the extensive dynamical evolution of the system, we find that the stellar multiplicity is almost constant during the 10~Myr of evolution.  This is because the primordial multiple systems are formed in a clustered environment and, thus, by their nature are already resistant to further evolution.  The majority of multiple system evolution is confined to the decay of high-order systems (particularly quadruple systems) and the formation of a significant population of very wide ($10^4-10^5$~AU) multiple systems in the expanding halo. This formation mechanism for wide binaries potentially solves the problem of how most stars apparently form in clusters and yet a substantial population of wide binaries exist in the field.  We also find that many of these wide binaries and the binaries produced by the decay of high-order multiple systems have unequal mass components, potentially solving the problem that hydrodynamical simulations of star formation are found to under-produce unequal-mass solar-type binaries.
\end{abstract}


\section{Introduction}
\label{introduction}

For many decades, it has been possible to perform {\it N}-body simulations of the evolution of stellar clusters, examining their dynamical evolution and the evolution of their stellar populations \citep{vonHoerner1960,vanAlbada1968a,AarWoo1972}.  Some early work was even done on the dynamical evolution of young clusters. For example, \cite{AarHil1972} studied the dynamical evolution of a stellar cluster with initial sub-clustering and found that sub-clustering increased both the fraction of stars ejected during the relaxation of the cluster and the formation of binary stars.  However, in the last two decades, during which observational studies of star-forming regions have dramatically increased in number, much more attention has been paid to the dynamical evolution of young stellar clusters.

\cite{Kroupa1995a} hypothesised that all stars were formed in binary systems and studied whether such initial conditions could lead to the observed field population of binaries.  With an initial period, $P$, distribution that was flat in $\log P$ ranging from $10^3$ to $10^{7.5}$ days, \citeauthor{Kroupa1995a} found that subsequent dynamical interactions in dense stellar clusters could reduce the binary fraction from 100\% and modified the period distribution to match the properties found for field stars.  On the other hand, simulations that begin only with binaries cannot reproduce the observed frequencies or properties of triples \citep{Portegiesetal2004} and simulations that begin with primordial binaries and triples cannot reproduce the observed frequencies of quadruples \citep*{vanPorMcM2007}.  From his simulations of modest-sized clusters containing 200 binaries, \cite{Kroupa1995b} also found that the clusters would be surrounded by a halo of ejected stars with a lower binary fraction. \cite{Kroupa1998} found that ejected binary systems tended to be massive with a preference for equal-mass components.  He also found that the mean system mass was approximately independent of ejection velocity from 2-30~km~s$^{-1}$, with dynamical ejection providing relatively massive stars as far as 300 pc from the cluster within 10 Myr \citep[see also][]{SteDur1995}.  The binary fraction decreased monotonically with ejection velocity for low-density clusters, but for dense clusters the dependence was more complex.  However, long-period systems (periods $>10^4$ days) could not be ejected with large velocities ($> 30$~km~s$^{-1}$).

More recently, attention has turned to the effects of gas dispersal on the evolution of young clusters and, following the pioneering work of \cite{AarHil1972}, to the effects of sub-structure.  Simple analytical models indicate that a young cluster requires a star formation efficiency of $\eta>50$\% to survive the rapid dispersal of gas as a bound cluster \citep{Hills1980}.  However, more sophisticated semi-analytic and numerical models show that the survival of a bound cluster is more complex.  If the gas is removed on a timescale significantly longer than the crossing time, then $\eta$ as low as 20\% can still result in a bound cluster \citep{Elmegreen1983,Mathieu1983,VerDav1989,GeyBur2001,BauKro2007} and even if the gas is removed quickly many stars can be lost, but a bound core can remain \citep*{LadMarDea1984,Goodwin1997,Adams2000,GeyBur2001,KroAarHur2001,BoiKro2003a,BoiKro2003b,BauKro2007}.  For example, \cite{GeyBur2001} and \cite{BoiKro2003b} find that with $\eta=0.4$ and rapid explusion around 30-40\% of the initial cluster could remain bound.   \cite{BauKro2007} find that for instantaneous gas removal $\eta>1/3$ is required for such a remnant cluster to be formed, but that if the gas is removed over several crossing times, around 50\% of the cluster can remain bound even for $\eta$ as low as 15-20\%.  Finally, \cite{Goodwin2009} notes that the survivability of a cluster depends not only on the star formation efficiency, but on the cluster's dynamical state.  In particular, if the gas removal occurs when the cluster is still dynamically cool (subvirial), then cluster survivability can be greatly increased.

The dynamical state of a cluster is also important for mass segregation. \cite{BonDav1998} showed that the Orion Nebula Cluster (ONC) is not old enough for the observed segregation of its massive stars to be due to dynamical segregation alone.  Instead, they argued that primordial segregation was required.  Recently, however, \cite{McMVesPor2007}, \cite{FelWilKro2009}, and \cite{MoeBon2009} have shown that if the cluster is formed via the merging of sub-clusters, mass segregation can be accelerated.  Similarly, \cite{Allisonetal2009a} investigated the evolution of dynamically cool clusters with fractal initial structure and found that dynamical mass segregation of the most massive stars can occur quickly.

However, the above studies have begun with artificial initial conditions.  Assumptions must be made about the initial structure of the stellar clusters (e.g. Plummer spheres, fractal or sub-clustered stellar distributions), the virial ratio (e.g. in virial equilibrium, sub-virial so they collapse, or super-virial so they expand), the initial populations of binary (e.g. 100\% binaries) and higher-order multiple systems (typically, no triples or higher-order systems, with the exception of \citealt{vanPorMcM2007}), and the properties of these multiples such as their distributions of separations and mass ratios.

Recently, it has become possible to perform hydrodynamical simulations of the collapse of a molecular cloud to form a star cluster containing in excess of 1000 of stars and brown dwarfs while resolving the opacity limit for fragmentation \citep{Bate2009a}.  Such calculations allow detailed comparisons to be made with the observed properties of young clusters, such as the stellar initial mass function (IMF), multiplicity, the properties of binary and multiple systems, and the global properties of stellar clusters.  Bate's calculation included gravity and hydrodynamics using a barotropic equation of state and used sink particles to model the stars and brown dwarfs.  The sink particles had accretion radii of only 5 AU, allowing circumstellar discs to be resolved down to $\approx 10$AU in radius, while binaries and multiple systems with separations as close as 1 AU could be followed.  Despite the limited amount of physics included in the calculations, many of the stellar properties obtained from the calculation were in good agreement with the observed properties of stellar systems.  For example, the observed stellar multiplicity as a function of primary mass was found to be well reproduced, as were the separations and mass ratios of low-mass stars and brown dwarfs.  Even the distribution of the orientation of the orbital planes of triple stellar systems was found to be in good agreement with observations.  The two main deficiencies of the calculation were that it produced a ratio of brown dwarfs to stars well in excess of that which is observed, and solar-type binaries were found to have a deficit of low mass ratio systems.  The first of these is likely to be due to the neglect of radiative feedback from the protostars \citep{Bate2009b,Offneretal2009}, while the solution to the latter problem is currently unclear.  However, the implication of Bate's calculation is that many stellar properties result primarily from dissipative gravitational dynamics and that additional physical processes such as radiative transfer and magnetic fields may not be very important for determining these properties.

One objection that could be raised to \cite{Bate2009a} is that, due to computational limitations, the formation of the cluster could only be followed for $2.85\times 10^5$~yr (1.5 global free-fall times) which was only $ 1.5\times 10^5$~yr after the first star formed. However, in many cases, the stellar properties were compared to star-forming regions that were typically $\sim 1$~Myr old, or even to field stars.  When Bate's calculation was stopped, there were many triple, quadruple, and higher-order stellar systems, and many of there systems were dynamically unstable.  Furthermore, of the 500~M$_\odot$ of gas originally contained in the molecular cloud only 191~M$_\odot$ (38\%) had been converted to stars by the end of the calculation.  Star formation would have continued if the calculation were continued.  But even if the gas was dispersed at this point and star formation ceased, there is the question of how the stellar properties would evolve over million-year timescales and longer.

In this paper, we take the end point of the hydrodynamical simulation of \cite{Bate2009a} as the starting point for {\it N}-body simulations.  The advantage of these starting conditions is that the stellar cluster has been formed self-consistently from the evolution of a gravitationally-unstable turbulent molecular cloud.  Thus, the structure of the cluster and the properties of binary and multiple systems have all been consistently set up, albeit with a restricted number of physical processes (e.g. no radiative feedback or magnetic fields).  Our aims are to determine the long-term (10~Myr) evolution of the cluster and how its properties and those of the stellar multiple systems evolve during the dispersal of the gas in the cluster.  The only previous similar work we are aware of is that of \cite{HurBek2008} who also performed SPH simulations of star formation to generate the initial conditions for {\it N}-body simulations.  However, their work studied star cluster formation on giant molecular cloud scales, assumed equal-mass stars, and did not resolve primordial multiple systems.

The paper is structured as follows.  Section 2 briefly describes the initial conditions and the numerical method for the calculations.  The results are discussed in Section 3.  In Section 4, we discuss the implications of the results for our understanding of star cluster evolution.  Our conclusions are given in Section 5.

\section{Computational method}

The calculations discussed in this paper are continuations of a hydrodynamical calculation that began with a turbulent molecular cloud and produced a star cluster.  The hydrodynamical calculation was performed using a smoothed particle hydrodynamics (SPH) code \citep*{BatBonPri1995} and the cloud was evolved for 1.5 initial cloud free-fall times ($2.85\times 10^5$~yr, where the free-fall time was $t_{\rm ff}=6.0\times 10^{12}$~s or $1.90\times 10^5$~yr).  The stars and brown dwarfs were modelled using sink particles.  In this paper, we take the sink particles at the end point of the hydrodynamical calculation as the starting point for a series of {\it N}-body simulations and evolve the stellar cluster to an age of 10 Myr.  Note that for all the times we quote, the starting point is taken to be that of the hydrodynamical calculation, not the {\it N}-body simulations, unless stated otherwise.

\subsection{Initial conditions}
\label{initialcond}

The initial conditions for the {\it N}-body simulations are taken from the main star cluster formation calculation of \cite{Bate2009a}.  The initial conditions for the hydrodynamical calculation consisted of a 500~M$_\odot$, uniform, spherical molecular cloud represented by $3.5\times 10^7$ SPH particles. The cloud's radius was 0.404 pc (83 300 au). At the initial temperature of 10~K, the mean thermal Jeans mass was 1~M$_\odot$ (i.e. the cloud contained 500 thermal Jeans masses). Although the cloud was uniform in density, an initial supersonic ÔturbulentÕ velocity field was imposed in the same manner as \cite*{OstStoGam2001}. A divergence-free random Gaussian velocity field was generated with a power spectrum $P(k)\propto k^{-4}$,where $k$ is the wavenumber. In three dimensions, this results in a velocity dispersion that varies with distance, $\lambda$,as $\sigma(\lambda)\propto \lambda^{1/2}$, in agreement with the observed Larson scaling relations for molecular clouds \citep{Larson1981}. The velocity field was normalized so that the kinetic energy of the turbulence equaled the magnitude of the gravitational potential energy of the cloud. Thus, the initial rms Mach number of the turbulence was $M=13.7$.  The `turbulence' decayed as the calculation was evolved; there was no `turbulent driving'.  The cloud was allowed to evolve freely into the vacuum surrounding it; there were no boundary conditions applied to the 
simulation.

\subsubsection{Sink particles}
\label{sinkparticles}

To model the thermal behaviour of the gas without performing radiative transfer, the SPH calculation used a barotropic equation of state (see \cite{BatBonBro2003} for further details).  This equation of state kept the temperature at 10~K for densities $<10^{-13}$ g~cm$^{-3}$, while above this density the temperature increased with density in order to mimic the  opacity limit for fragmentation \citep{LowLyn1976,Rees1976,Silk1977a,Silk1977b,BoyWhi2005}.  When the central density of a fragment exceeded 
$\rho_{\rm s} = 10^{-11}~{\rm g~cm}^{-3}$, 
it was replaced by a sink particle \citep{BatBonPri1995}.  In the main calculation of \cite{Bate2009a}, sink particles were formed by 
replacing the SPH gas particles contained within $r_{\rm acc}=5$ AU 
of the densest gas particle in a pressure-supported fragment 
by a point mass with the same mass and momentum.  Any gas that 
later fell within this radius was accreted by the point mass 
if it was bound and its specific angular momentum was less than 
that required to form a circular orbit at radius $r_{\rm acc}$ 
from the sink particle. Sink particles interacted with the gas only via gravity and accretion. The gravitational acceleration between two sink particles was Newtonian for $r\geq 4$ AU, but was softened within this radius
using spline softening \cite{Benz1990}.  The maximum acceleration 
occurs at a distance of $\approx 1$ AU; therefore, this is the
minimum separation that a binary had in the hydrodynamical simulation even if, in reality, a binary's orbit would have been hardened.  Sink particles were permitted to merge if they passed within $2R_\odot$ of each other; only one merger occurred during the simulation.

\subsubsection{The end state of the hydrodynamical calculation}

The end state of the hydrodynamical calculation consisted of a cluster of 1253 stars and brown dwarfs with a combined mass of 191~M$_\odot$ while 309 M$_\odot$ of gas remained.  Note that \cite{Bate2009a} states that there were 1254 objects, but in fact it was not noticed when the paper was written that two of these had merged and had been replaced by a single object.  This has no significant effect on the statistics that were reported in the original paper.  The final cluster was created from the merger of around half a dozen sub-clusters that formed in gaseous filaments that were produced in the turbulent gas.  The sub-clusters formed in a dynamically cool configuration and fell together to produce a single cluster surrounded by an extended halo of unbound, dynamically ejected stars and brown dwarfs.  This cluster was very dense, with a half-mass radius of only $10^4$ AU (0.05~pc) and a velocity dispersion of around 4 km~s$^{-1}$, with the velocities of the ejected halo objects extending beyond 20~km~s$^{-1}$.  Many of the stars were in binary or higher-order multiple systems and the stellar multiplicity was found to be a strongly increasing function of primary mass, with values in good agreement with observed values. Many of the trends for the separation and mass ratio distributions of observed binaries were also reproduced (e.g. the trend for very-low-mass binaries to have smaller separations and mass ratios nearer unity than for low-mass and solar-type stars), with the exception that unequal-mass solar-type binaries were under-produced.

Further details regarding the initial conditions, the SPH code, the evolution of the hydrodynamical star cluster formation calculation and the statistical properties of the stars and brown dwarfs can be found in \cite{Bate2009a}.  We now turn to the question of how this initial cluster evolves on 10~Myr timescales.

\begin{figure}
 \includegraphics[width=8.0cm]{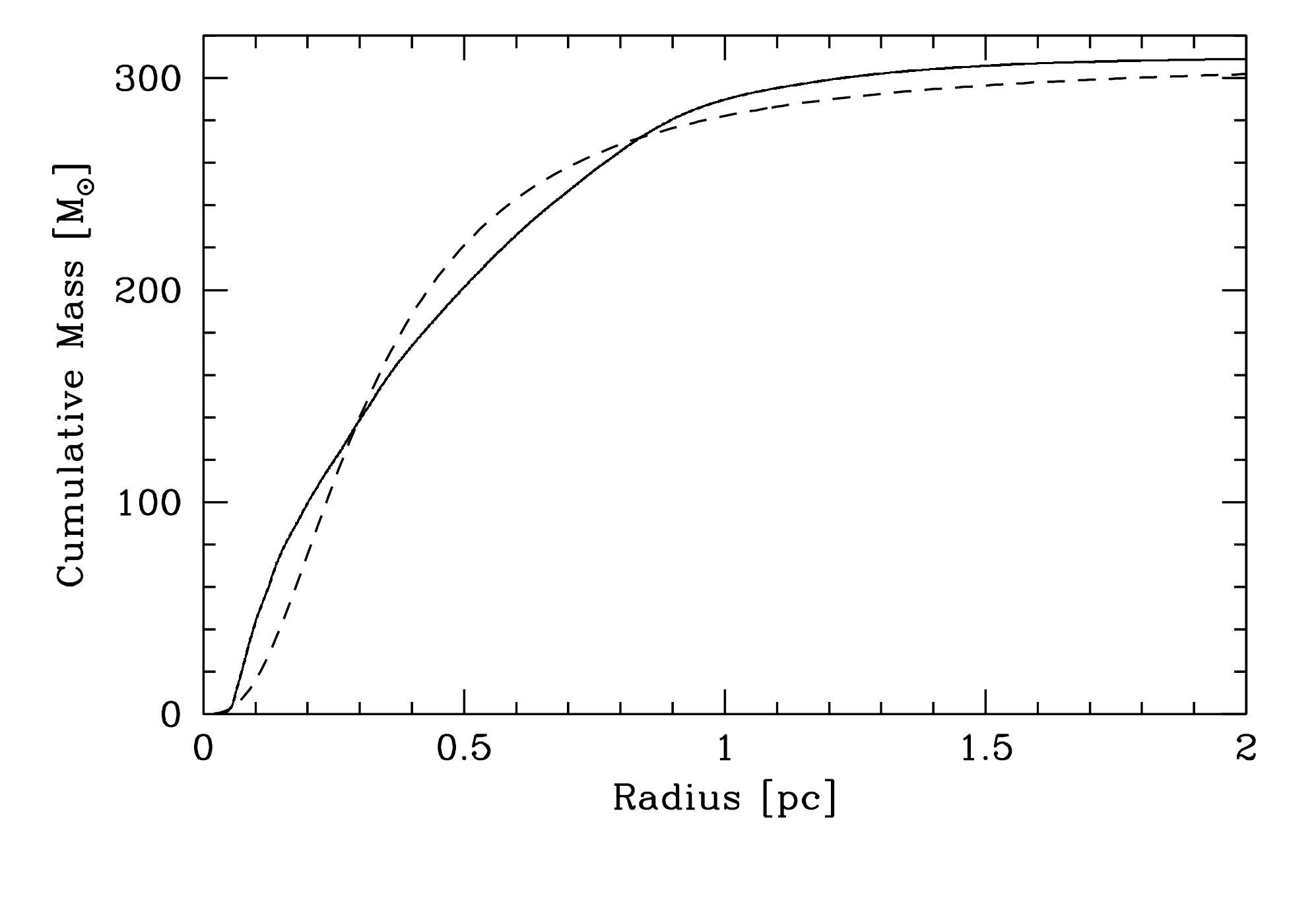}
 \caption{The solid line gives the cumulative radial distribution of the 309~M$_\odot$ of gas that remained at the end of the hydrodynamical calculation of Bate (2009a), although the gas distribution is not spherically symmetric.  Note that the half-mass radius of the stellar cluster is only 0.05 pc.  The dotted line gives the cumulative radial distribution of a Plummer sphere with the same total mass and a characteristic radius of 0.25~pc.  The two distributions have similar half-mass radii.}
 \label{gasfig}
\end{figure}

\subsection{The {\it N}-body calculations}

\begin{table*}
\begin{center}
\begin{tabular}{lcccccccc}\hline
Calculation & End Time & \multicolumn{2}{c}{Remnant Cluster} & Mass & \multicolumn{4}{c}{Final Numbers of Systems}  \\
& [Myr] & Mass & Radius [pc] & Segregation & Single & Binary  & Triple & Quadruple  \\ \hline
Hydrodynamical (Bate 2009a)      & 0.3 & -- & -- & $\approx 10$ most masssive stars &      905      &     90     &      23      &     25    \\
Instant Gas Removal     & 10 & 30\% & 2 &  -- &   956      &     98    &       23    &      8   \\
$T_{0.1}=0.3$~Myr      & 10 & 30\% & 1.8 & $\gsim 1$~M$_\odot$  &  941      &   100    &       20     &      13   \\
$T_{0.1}=1$~Myr    & 10 & 40\% & 1.3 & $\gsim 1$~M$_\odot$  &      962      &    97      &     19     &      10   \\
No Gas Removal     & 10 & 50\% & 0.8 & $\gsim 0.3$~M$_\odot$  &      966     &     93    &       11     &      17   \\ \hline
\end{tabular}
\end{center}
\caption{\label{summarytable} A summary of the calculations discussed in this paper.  The four {\it N}-body calculations with different gas removal timescales take their initial conditions as the stars (sink particles) at end of the hydrodynamical calculation of \citet{Bate2009a}.  $T_{0.1}$ is the time from the start of the {\it N}-body evolution at which the gas mass has fallen to 10\% of its initial value.  In addition to the calculations listed above, the instantaneous gas removal case was run ten times with randomly perturbed initial conditions.  In the table, we summarise the properties of the remnant clusters that are left at the end (10~Myr) of the {\it N}-body calculations (their radii and the mass fractions of the stars they contain).  We also comment on the mass segregation that is observed (e.g. stars with masses $>0.3$~M$_\odot$ at the end of the case with no gas removal).  Finally, we give the numbers of single and multiple systems at the end of the hydrodynamical calculation of \citet{Bate2009a} and each of the unperturbed {\it N}-body calculations.  The {\it N}-body evolution always results in a significant decrease in the number of quadruple systems and small increases in the numbers of single and binary systems.  Slower gas removal timescales generally produce more compact and populous remnant clusters that show more mass segregation.}
\end{table*}

The final positions, masses and velocities of the sink particles in the hydrodynamic simulation are taken as the initial conditions for a gravitational {\it N}-body simulation.  The integration was performed with the {\sc nbody6} code \citep{Aarseth2003}, which uses a fourth-order Hermite integrator.  We advanced the system for 10 Myr from the end of the hydrodynamic simulation, and the relative energy error was of order $10^{-5}$.  It is usual practice when performing {\it N}-body calculations with a modest number of stars to generate many sets of initial conditions and consider ensemble quantities, so that general trends can be separated from chaotic variations in the integrations.  In this case, with initial conditions generated not by random realizations of assumed bulk cluster properties but by a self-consistent and expensive hydrodynamical calculation, we are forced to take a slightly different approach to disentangling general results from those particular to the exact initial conditions.

Ten sets of perturbed initial conditions were generated by taking the initial conditions and randomising each body's position by $10^{-4}$ times its radius from the cluster's center of mass.  That is, each position vector ${\bf r}$ was adjust so that ${\bf r} \rightarrow {\bf r} + 10^{-4}r {\bf w}$, with ${\bf w}$ a randomly oriented unit vector.  The initial specific potential energy of each body was thus perturbed by an amount greater than the integration accuracy.  Stars in binaries were treated as single bodies for the perturbation, so that their initial binary properties remained the same.  By comparing results of the main integration to those of the perturbed versions of the initial conditions, we can ascertain which long-term results are robust and general to the type of cluster generated by the hydrodynamic simulation, and which are sensitive to the exact state of the cluster.  

The main calculation and the ten randomly perturbed calculations are all run assuming that the gas left at the end of the hydrodynamical calculation is instantaneously removed.  This is the most extreme gas removal history possible and, based on previous work, should result in the least-bound cluster remnant possible \citep{LadMarDea1984,Goodwin1997,Adams2000,GeyBur2001,KroAarHur2001,
BoiKro2003a,BoiKro2003b,BauKro2007}.  If the gas is removed over a finite period of time, we would expect more of the cluster to remain bound and it might also affect the frequencies and properties of the multiple stellar systems.  To investigate how much of a difference slow removal of the gas makes, we have performed three additional calculations that include an additional gravitational potential to model the contribution of the gas.

The gas mass at the end of the hydrodynamic calculation is 309~M$_\odot$. While the actual gas distribution is not spherical \citep[see Figure 1 of ][]{Bate2009a}, we approximate it as a Plummer sphere with scale radius 0.25 pc so that the half-mass radius is similar to that of the actual gas distribution (see Figure \ref{gasfig}).  The mass of the Plummer potential is decreased with time, $t$, from the beginning of the {\it N}-body calculations so that the gas mass is given by
\begin{equation}
  M_{\rm gas} = \frac{M_{0}}{1 + t/\tau_{\rm gas}},
\end{equation}
where $M_{0}$ is the original gas mass (309~M$_\odot$) and $\tau_{gas}$ is a gas removal timescale.  We ran three cases: moderately fast gas dispersal, where the mass was reduced to 10\% of its initial value after 0.3 Myr of {\it N}-body evolution (total age 0.6~Myr); slow dispersal, where the mass was reduced to 10\% at 1 Myr  of {\it N}-body evolution (total age 1.3~Myr); and an (unphysical) test case where the gas mass remained constant throughout the calculation.

A summary of the main properties of the calculations discussed in the next section is presented in Table \ref{summarytable}. The plots and discussion that follow are based on the unperturbed initial conditions with instantaneous gas removal, with comparison to the randomized integrations and those with gas potentials made where appropriate.

\begin{figure*}
 \includegraphics[width=13.5cm]{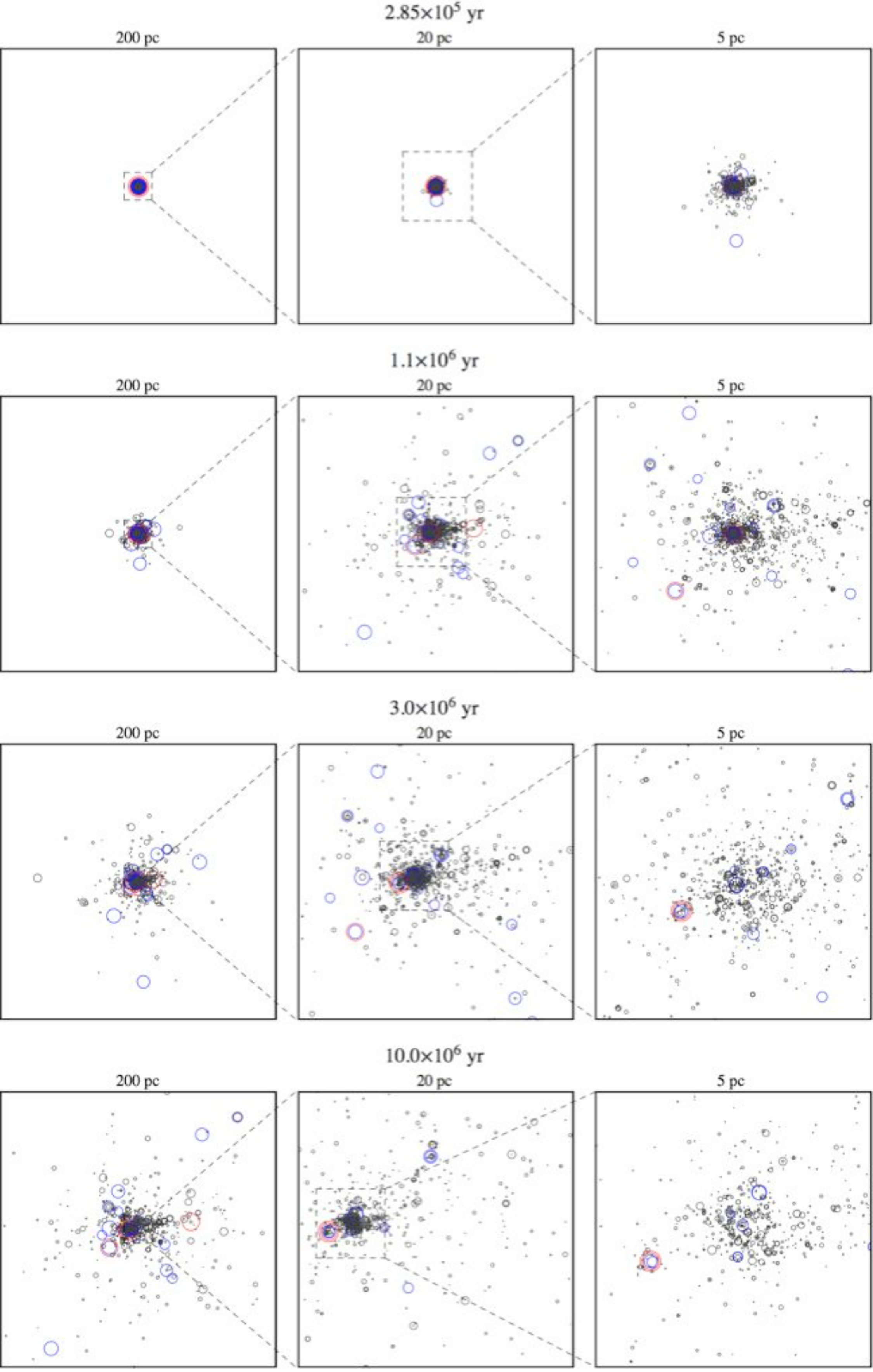}
 \caption{The {\it N}-body evolution of the unperturbed star cluster with instantaneous gas removal.  Each star is plotted using an open circle with the area of the circle proportional to the star's mass.  Stars with masses $\ge 3$~M$_\odot$ are plotted in red, while those with masses $1 \le M < 3$~M$_\odot$ are plotted in blue.  Binaries and multiples can be seen as almost concentric circles when the components have different masses.  We show the structure on 200, 20, and 5~pc scales at four different times.  On the largest scales, it can be seen that some ejected stars in the halo reach distances $>100$~pc in less than 10~Myr and that these stars are a mixture of massive and low-mass stars.  On the smallest scales, the remnant bound cluster containing approximately 1/3 of the stars expands from its initial radius of $\approx 0.05$~pc to $\approx 2$~pc.  The most massive star and a small cluster of $\approx 20$ companion stars is ejected from the remnant cluster at $\approx 1$~Myr and can be seen to the lower-left of the main cluster remnant in the lower two 5~pc panels.}
 \label{ClusterOverview}
\end{figure*}

\section{Results}
\label{results}

\subsection{Evolution of cluster properties}

\subsubsection{Substructure}

In the hydrodynamic simulation of \citet{Bate2009a}, the stars form in a structured fashion, with smaller subclusters merging to form a final cluster consisting of a tightly bound core with radius $\approx 0.05$~pc surrounded by an expanding halo of ejected stars. The reader interested in the hydrodynamical evolution of the cluster is encouraged to consult Figure 1 of \cite{Bate2009a}. In Figure \ref{ClusterOverview} of this paper, we display the global evolution of the stellar cluster during the 10~Myr of {\it N}-body evolution.  Each star is represented by an open circle with its area proportional to the star's mass.  On large scales, the stars in the halo are seen to expand to distances exceeding 100~pc from the cluster in less than 10~Myr.  On small scales, a bound cluster remnant containing approximately 30\% of the stars expands from a radius of $\approx 0.05$~pc to $\approx 2$~pc over several Myr.

\begin{figure}
 \includegraphics[width=80mm]{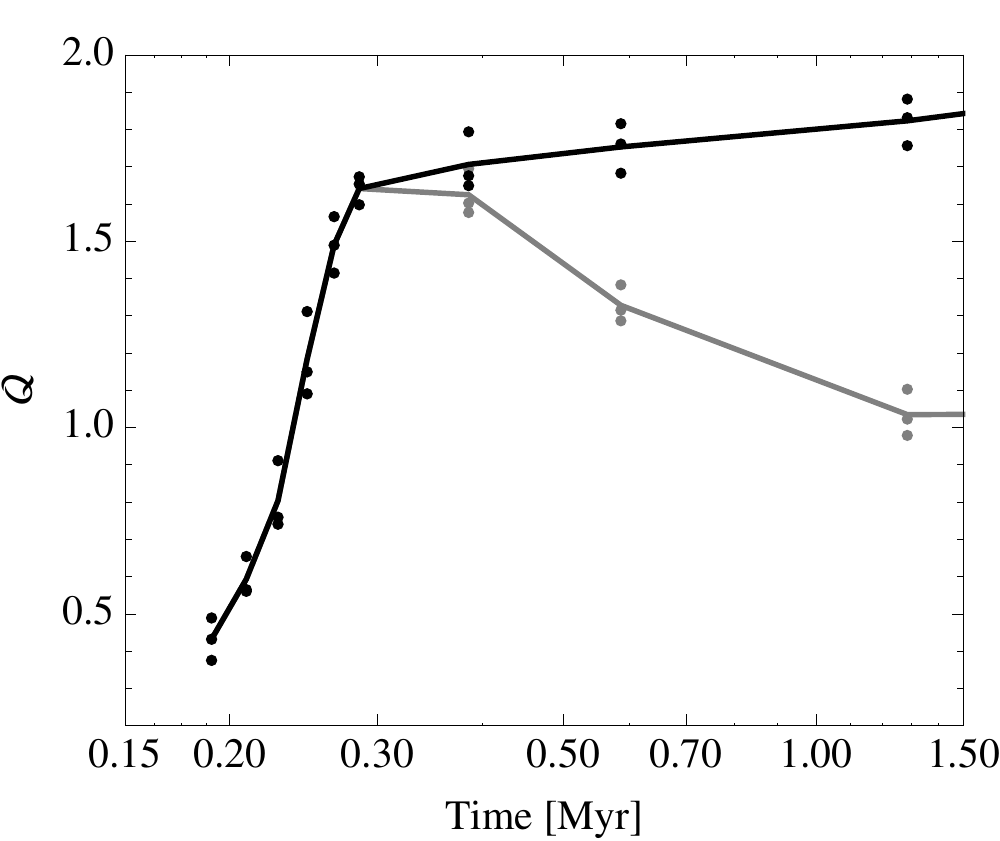}
 \caption{The value of the parameter $Q$ over the first 1.5 Myr.  At each time the quantity is determined for three orthogonal projections of the cluster, shown as the points, and the means of those values are connected by the lines.  The black line gives the evolution of the entire stellar cluster and halo, while the grey line gives the evolution of the inner 3~pc region.  The inner region evolves from a sub-clustered stellar distribution ($Q<0.8$) to a core and halo structure ($Q>1$) during the hydrodynamical calculation.  When the gas is removed, the remnant core expands to fill most of the inner 3-pc and $Q$ decreases to approximately unity.}
 \label{Q_evolution}
\end{figure}

The degree of subclustering can be quantified by the parameter $Q$ \citep{CarWhi2004}.  Values of $Q < 0.8$ indicate substructure and can be associated with a fractal dimension, while values greater than 0.8 are associated with radial density variation.  Calculating $Q$ requires both an area and radius to be assigned to a cluster; as the cluster at early times is decidedly non-spherical, we calculate the area as the convex hull of the stars' positions projected onto a plane, and take the radius to be that of a circle having the same area.  The evolution of $Q$ over the first 1.5 Myr of the cluster evolution is shown in Figure \ref{Q_evolution}.

At the earliest measured times the cluster consists of 154 stars, and the value of $Q \approx 0.4$ corresponds to an estimated fractal dimension of  1.5, i.e. highly substructured.  Over the next $\sim 4\times10^4$ yr the early subclusters merge, the number of stars increases to 579, and $Q$ rises above 0.8.  This indicates a fractal dimension of 3, i.e. a distribution lacking any significant substructure.  By the end of the hydrodynamical calculation the number of stars has reached its final value of 1253, and $Q$ has risen in excess of 1.5.  In \citet{CarWhi2004}, larger values of $Q$ are identified as corresponding to an increasingly steep radial density profile of a spherical cluster.  In our case the cluster is not well-fit by a single power law density profile, but rather has a core-halo structure, which complicates a straightforward interpretation of $Q$.  Because the cluster is characterized by a bound remnant surrounded by an expanding halo, the exact value we calculate is quite sensitive to the radial extent of stars that we consider.  In Figure \ref{Q_evolution}, following gas removal, two lines are plotted.  The black line includes all the stars, arranged with a core-halo structure, while for the grey line the analysis has been restricted to the central 3-pc of the cluster.  In this central region, the cluster core expands upon gas removal to occupy most of the inner region and the value of $Q$ drops to unity, reflecting an absence of significant structure and the fact that the halo is only present on larger scales.  For our purposes it suffices to note that $Q$ remains quite high for the entire 10 Myr run, showing the persistence of the core-halo structure. 

We emphasise that the evolution of the cluster from a highly-clustered to a core-halo structure occurred while the the cluster was still embedded and forming; the number of stars increases by nearly an order of magnitude during the transition from a fractal dimension $\approx 1.5$ to $\approx 3.0$.  This highlights the importance of coupled gas and gravitational dynamics in young cluster evolution, and the difficulty in assigning realistic initial conditions to purely {\it N}-body studies.

\subsubsection{Lagrangian radii and stellar velocities}
\label{LagrangianSection}

\begin{figure}
 \includegraphics[width=84mm]{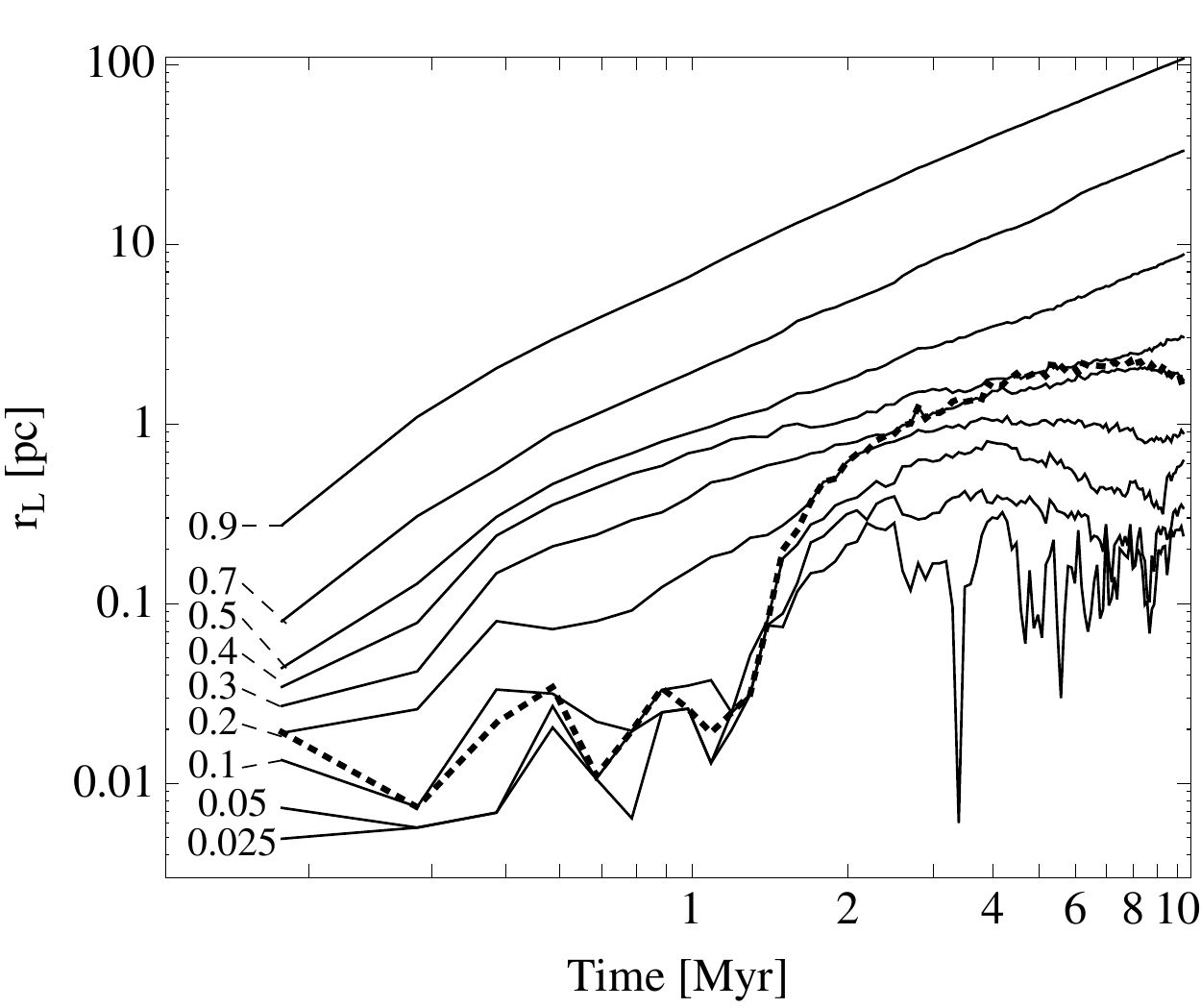}
 \caption{Lagrangian radii during the span of the main {\it N}-body calculation.  From bottom to top, solid curves enclose 0.025, 0.05, 0.1. 0.2, 0.3, 0.4, 0.5, 0.7, and 0.9 of the total stellar mass.  The dotted line shows the median radius of the five most massive stars.}
 \label{LagrangianRadii}
\end{figure}

The radial mass structure of the cluster can be seen in the evolution of the Lagrangian radii, that is the radii encompassing a constant percentage of the cluster's mass.  This calculation is dependent on what one considers the cluster center.  At each time we identify the star which is located in the region of the greatest stellar density, which we define as the star with the smallest sphere encompassing 50 neighbours, and set this star as the origin.  The behavior of these diagnostics depends sensitively on the details of the integration, especially the behavior of the most massive stars.  We first discuss the main features of the Lagrangian radii for the unperturbed initial conditions, shown in Figure \ref{LagrangianRadii}, and then note how these vary among the randomly perturbed clusters and those with a gas potential.

\begin{figure}
 \includegraphics[width=90mm]{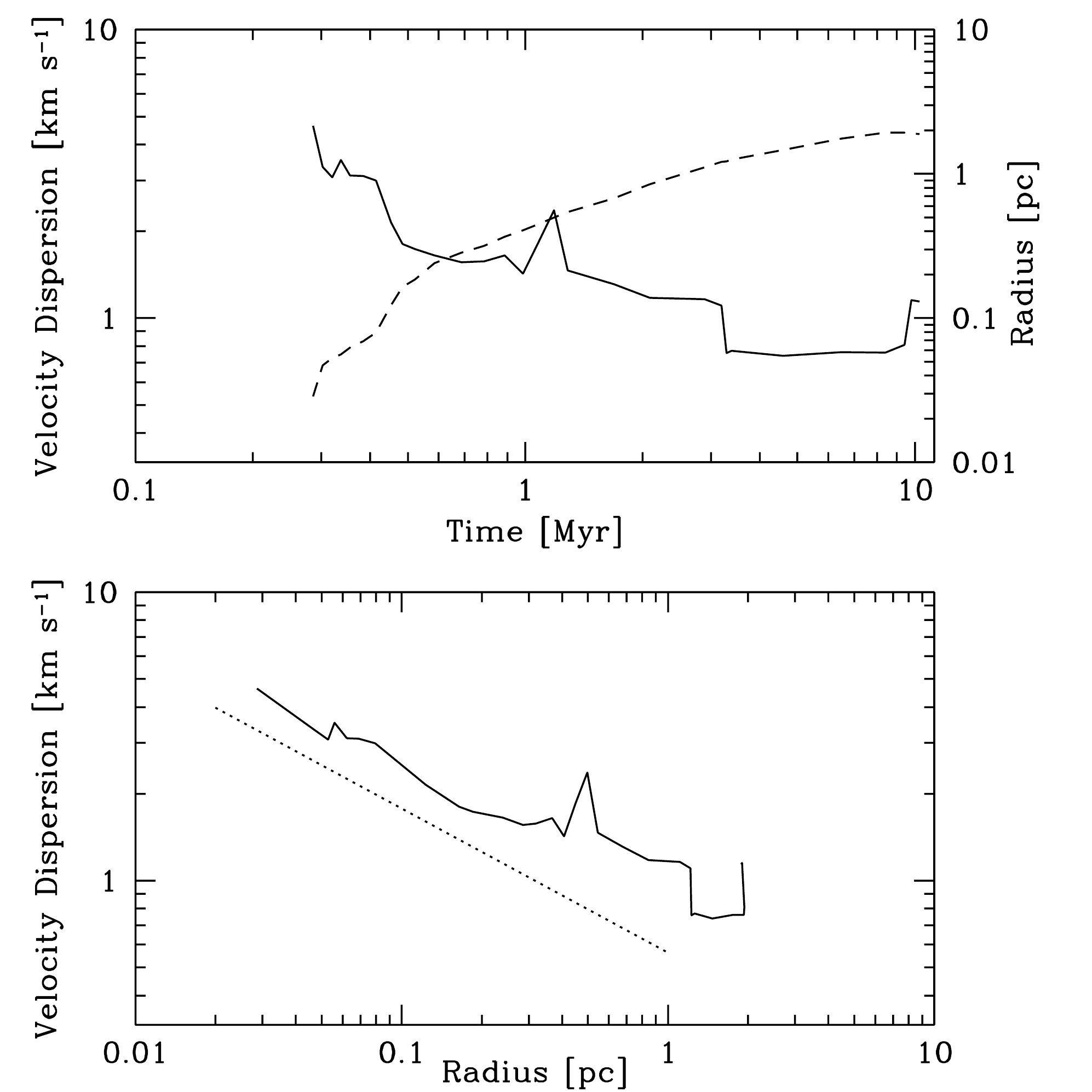}
 \caption{The evolution of the stellar velocity dispersion and the radius of the bound remnant cluster in the main {\it N}-body calculation.  In the top panel we give the time evolution of the radius containing 30\% of the stars (dashed line), centred on the star at the location of the highest stellar density.  It can be seen that the remnant cluster expands by a factor of $\approx 20$ in size.  The velocity dispersion of the stars in the remnant (taking the centre of mass velocity for all binaries and counting them only once) is given by the solid lines in both panels.  The velocity dispersion decreases from $\approx 4$ to $\approx 1$ km$^{-1}$.  The spike at $\approx 1$~Myr is associated with the ejection of the most massive binary from the cluster.  In the lower panel, the velocity dispersion is plotted as a function of the remnant cluster's radius.  The dotted line has a slope of $-1/2$, as expected if the remnant cluster remains close to virial equilibrium as it expands.}
 \label{velocity}
\end{figure}

The first Myr of {\it N}-body evolution is marked by general expansion of the outer 70\% of the cluster's mass, with the Lagrangian radii increasing by an order of magnitude.  In contrast, the inner 10\% of the mass expands by only a factor of 2.  Between 1 and 2 Myr these inner radii expand by an order of magnitude, catching up to the expansion of the outer radii.  At later times there is a divergence as the outer radii continue to expand, while the inner radii (certainly the inner 20\%, and after 8 Myr the inner 30\%) halt their growth, leaving a remnant cluster core at the center of an expanding halo.  The evolution of this remnant cluster core is further explored in Figure \ref{velocity} where we plot the velocity dispersion of the stars in the remnant as a function of both time and of the remnant cluster's radius.  As expected for a system near virial equilibrium, the velocities of the stars decrease as the remnant expands roughly as $R^{-1/2}$.

\begin{figure}
 \includegraphics[width=84mm]{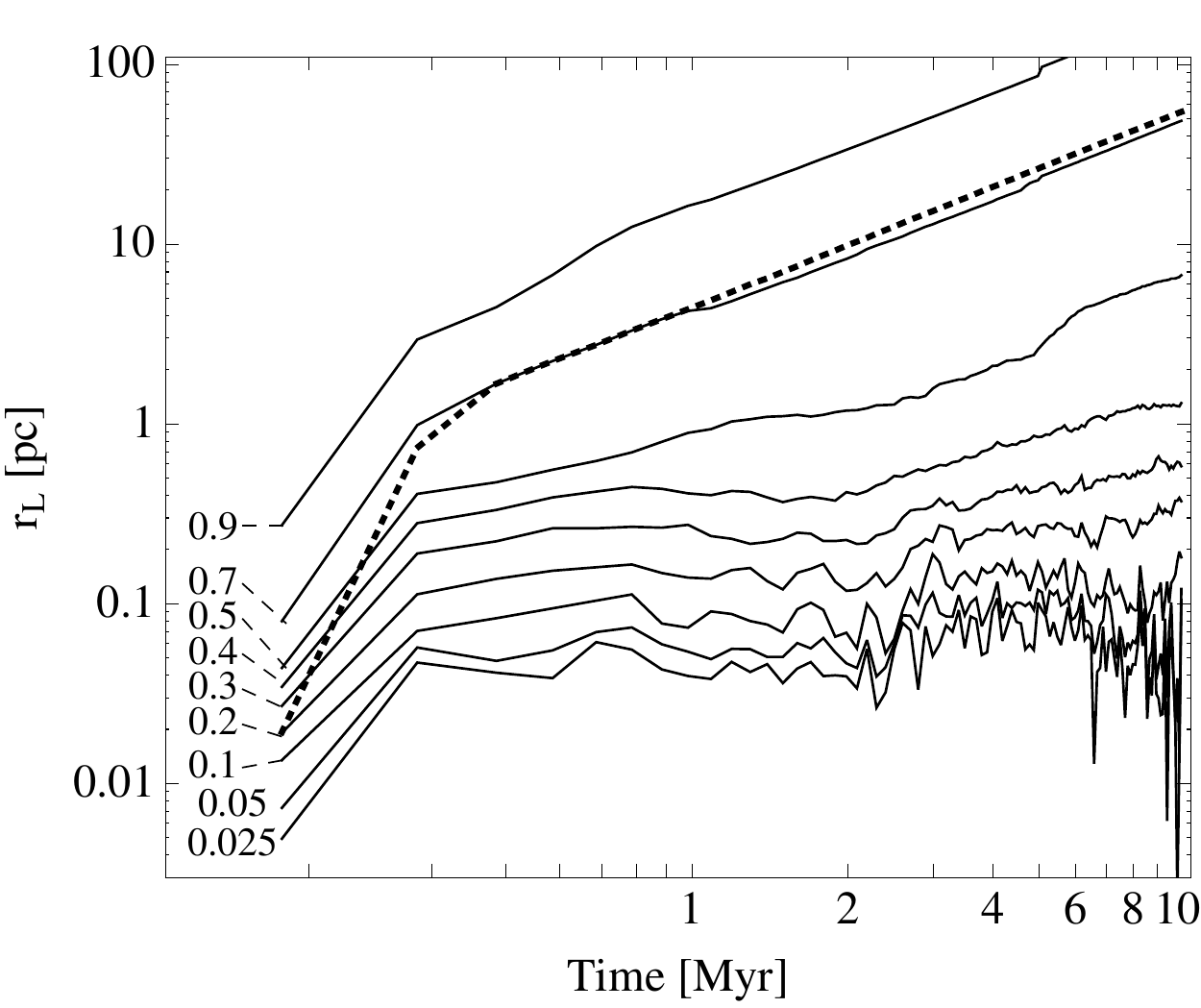}
 \caption{Lagrangian radii during the span of the {\it N}-body calculation that includes a gas potential which decreases to 10\% of its initial value over 1~Myr of evolution (total age 1.3~Myr).  From bottom to top, solid curves enclose 0.025, 0.05, 0.1. 0.2, 0.3, 0.4, 0.5, 0.7, and 0.9 of the total stellar mass.  The dotted line shows the median radius of the five most massive stars.  In comparison with Figure \ref{LagrangianRadii}, the remnant cluster is smaller and more populous, while more of the most massive stars have been ejected.}
 \label{1Myr_dispersal}
\end{figure}

The 1 Myr delay in the expansion of the inner 10\% of the mass is explained by the behavior of the most massive stars.  The 5 most massive stars in the cluster are by themselves approximately 10\% of the total stellar mass, and if they are concentrated in the cluster center then they will dominate the inner Lagrangian radii.  To give an idea of these stars' radii from the cluster centre, we also plot in Figure \ref{LagrangianRadii} the median radius of the 5 most massive stars as a dotted line.  At about 1 Myr a massive binary, comprised of a 5.35~M$_\odot$ primary (also the most massive star in the cluster) and a 3.54~M$_\odot$ secondary, is ejected from the cluster center (see the lower two right-hand panels of Figure \ref{ClusterOverview}).  This event also appears in the velocity dispersion evolution depicted in Figure \ref{velocity}.  The massive binary is accompanied by a small group of $\approx 20$ companion stars.  The inner 10\% Lagrangian radii track this binary and its companions for a few$\times 10^4$ years, at which point they begin to reflect the overall cluster structure rather than the positions of a few massive stars.  

This feature, a delayed expansion of the inner Lagrangian radii due to a clustering of the most massive stars, is seen in 9 of the 10 randomly perturbed simulations, though the time of the ejection ranges from $\sim 4\times 10^4$ yr to 2 Myr after the start of the {\it N}-body evolution.  In one case the massive stars remain in the cluster center throughout the 10 Myr.  In other words, in only one of our 11 runs  with instantaneous gas removal do the massive stars remain continually associated with the center of the cluster for more than 2 Myr.  In two cases the massive stars return to the center of the cluster, one of which sees a second ejection.  In that perturbed realization, after the second ejection of massive stars at about 3 Myr the inner Lagrange radii expand freely.  While a cluster remnant is still identifiable, it is globally dispersing unlike any of the other realizations.  Another feature is the behavior of the 20\% radius in the unperturbed case, which is fairly flat after its initial expansion.  In six of the perturbed runs this radius increases by a factor of 3 or less between 2 and 10 Myr; in two cases this radius is flat for most of the evolution but grows over the final 2 Myr; and in one case it expands freely after the ejection of the massive stars.  The global evolution of the cluster's mass profile thus appears to be surprisingly sensitive to the initial conditions.  We can say with some confidence that the massive stars are unlikely to remain associated with the main cluster remnant, but the final characteristics of that remnant (i.e. its radius, mass, and its stability or expansion) display significant variation.

\begin{figure}
 \includegraphics[width=84mm]{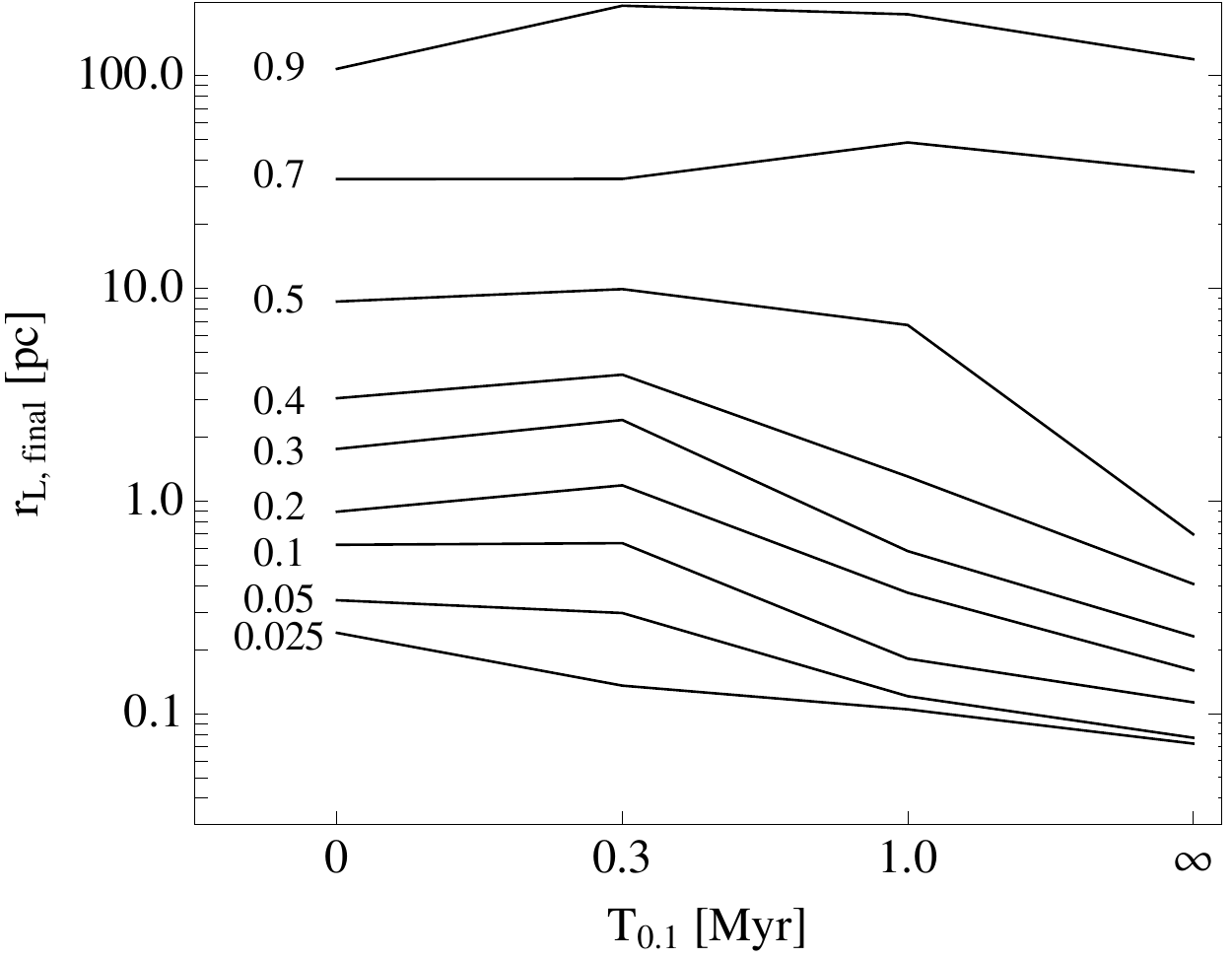}
 \caption{Lagrangian radii at the end of the {\it N}-body calculations (10~Myr) as a function of the gas removal timescale, $T_{0.1}$, which gives the time from the beginning of the {\it N}-body evolution at which the gas potential has been reduced to 10\% of its initial value (0, 0.3, 1~Myr and no gas removal).  Longer gas removal times yield more compact and populous clusters.}
 \label{DispersalVariation}
\end{figure}

As expected, the {\it N}-body calculations that include a gas potential and slower gas removal result in smaller final Lagrangian radii, and larger membership of the remnant cluster.    In Figure \ref{1Myr_dispersal}, we show the evolution of the Lagrangian radii in the slow gas dispersal case (where the gas mass is reduced to 10\% of its initial value after 1~Myr of ${\it N}$-body evolution).  Comparing this with Figure \ref{LagrangianRadii}, we find that somewhat more of the stellar mass (about 40\% compared to 30\%) remains in a remnant cluster and this remnant cluster is somewhat smaller in the slow gas dispersal case (radius approximation 1.3~pc rather than 2~pc).  The effect of the gas potential is more clearly demonstrated in Figure \ref{DispersalVariation}, where we show the Lagrangian radii at 10~Myr for each of the unperturbed cases, which we label by the timescale, $T_{0.1}$, for the gas mass to be reduced to 10\% of its initial value.  The inner Lagrangian radii all display the expected trend, with longer gas-removal times yielding a more compact cluster and populous cluster.  At 10 Myr the instantaneous, 0.3 Myr, 1.0 Myr, and no gas removal cases have 552, 595, 756, and 944 stars within 3 pc of the densest star, respectively.

\subsection{Mass segregation}

\begin{figure*}
\includegraphics[width=160mm]{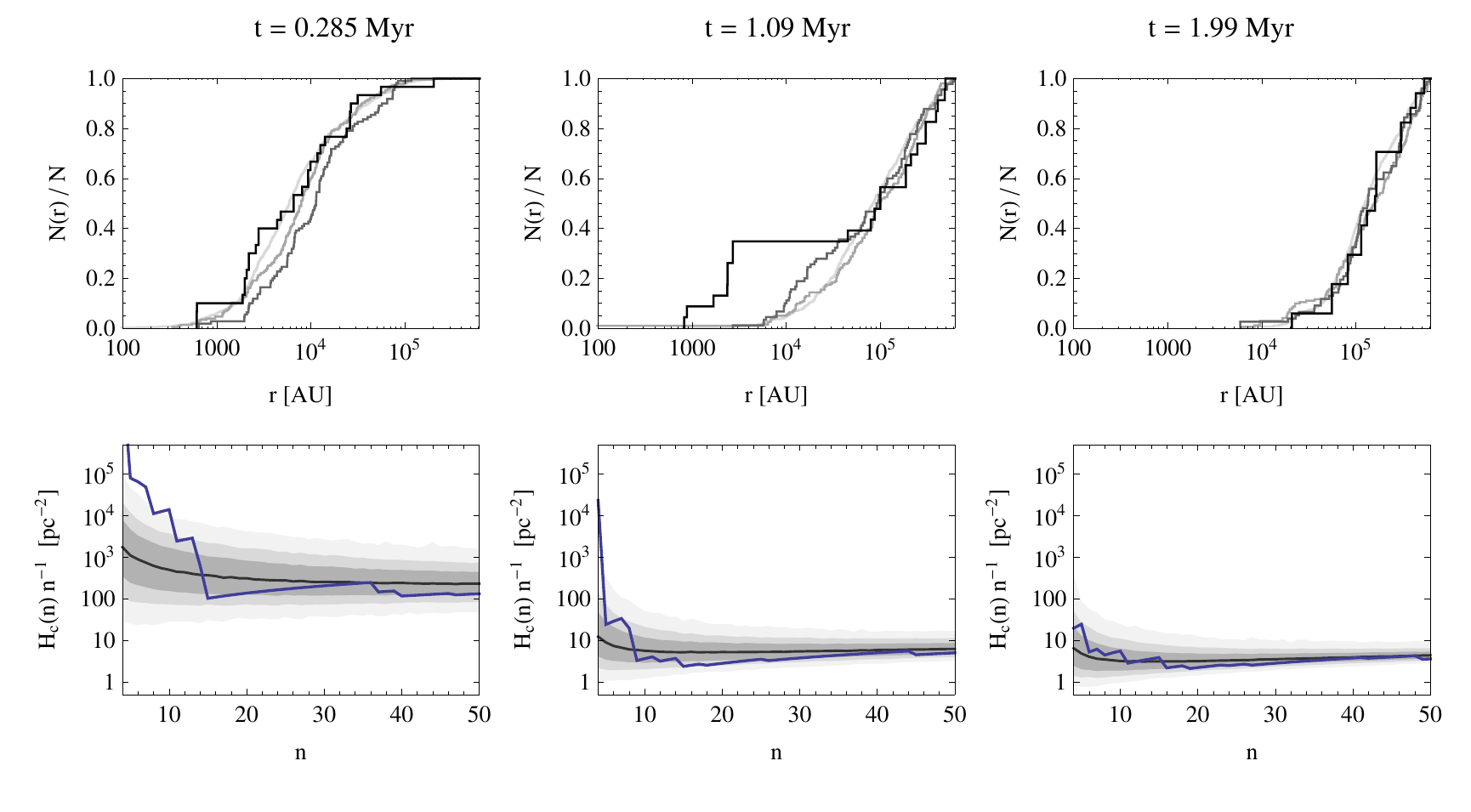}
 \caption{The mass segregation in the unperturbed cluster with instantaneous gas removal at three times; the end state of the hydrodynamic simulation, just before the ejection of the most massive stars from the cluster center, and at $\sim 2$ Myr.  {\it Top row:} segregation measured by the cumulative radial distribution of stars in different mass bins, measured from the position of the star in the region of highest stellar density.  The lines, from light grey to black, show stars in the mass bins $M/$\msun$< 0.1$, $0.1 \le M/$\msun$< 0.3$, $0.3 \le M/$\msun$<1.0$, $1.0 \le M/$\msun.  {\it Bottom row:} segregation measured by the effective surface density of the most massive $n$ stars (see the text for further explanation).  The gray line is the median value for the surface density of $n$ random stars, and the shaded regions show the $\pm1\sigma$, $\pm2\sigma$, and $\pm3\sigma$ values.  The blue line is the surface density of the $n$ most massive stars.
 }
 \label{MassSegregation}
\end{figure*}

Mass segregation is seen in several young clusters, typically appearing as a concentration of the most massive stars \citep[e.g. the ONC or Mon R2][]{HilHar1998, Carpenteretal1997} rather than a general trend affecting all mass ranges, as would be expected in older, dynamically mature clusters.  Recent work \citep{McMVesPor2007,FelWilKro2009,Allisonetal2009a,MoeBon2009} shows that a structured mode of star formation, with merging substructures forming the final quasi-spherical cluster, can lead to segregation affecting only the massive stars, mirroring observations.  \citet{MoeBon2009} noted that the final state of the \citet{Bate2009a} hydrodynamic simulation showed a similar mass segregation signal to the ONC.  Here, we examine how this rapid mass segregation evolves on longer timescales. 

With the realization that mass segregation in very young clusters is qualitatively different from the segregation that develops as a spherical cluster evolves over many relaxation times, new methods of quantifying segregation have been introduced.  We use two diagnostics here: first, a standard technique, the cumulative radial distribution of stars in different mass bins; second, a variation of the minimum spanning tree (MST) method introduced in \citet{Allisonetal2009b}, in which we measure the effective surface density of the most massive stars and compare it to the expected surface density.  This is the method used in \citet{MoeBon2009}, and full details are found there.  Briefly, the expected surface density of $n$ stars in the cluster is found by calculating the area of the convex hull, $H_c(n)$, for many samples of $n$ random stars.  The median, $\pm1\sigma$, $\pm2\sigma$, and $\pm3\sigma$ of the quantity $H_c(n)/n$ define the `normal' surface density of $n$ random stars.  This distribution is then compared to the surface density of the $n$ {\em most massive} stars.  With both of these methods, we restrict our analysis to stars within 3 pc of the star at the location of greatest stellar density.  This radius encompasses the final cluster remnant at 10 Myr, and prevents very distant escaping stars from interfering with a sensible interpretation of the results.

\begin{table*}
\begin{tabular}{lcccccc}\hline
Quantity /  Distance range & $<1$ pc & $1-3$ pc  & $3-10$ pc  &  $10-30$ pc & $30-100$ pc & $>100$  pc \\ \hline

Median mass [M$_\odot$] & 0.046 & 0.033 & 0.035 & 0.047 & 0.046 & 0.050 \\
Upper quartile mass [M$_\odot$]  & 0.14 & 0.09 & 0.11 & 0.14 & 0.15 & 0.21 \\
Maximum mass [M$_\odot$] & 2.9 & 5.3 & 2.3 & 1.7 & 3.7 & 3.7 \\
Velocity dispersion [km/s] &    0.8 & 1.3 & 0.7 & 2.1 & 5.7 & 17 \\
Number objects &    276 & 236 & 223 & 276 & 169 & 73 \\
Number binaries &  39 & 20 &  27 & 32 & 10 & 4 \\
Binary fraction &   0.16 & 0.09 & 0.14 & 0.13 & 0.06 & 0.06 \\\hline
\end{tabular}
\caption{\label{tablecluster} Radial properties of the remnant cluster (radius $\approx 2$~pc) and stellar halo (extending beyond 200~pc) at the end of the unperturbed {\it N}-body evolution with instantaneous gas removal (age 10~Myr).  This table can be compared with Table 2 of Bate (2009a) for the properties of the cluster at the end of the hydrodynamical evolution.  There is no evidence for radial mass segregation of the cluster in terms of the median, upper quartile, or maximum masses.  Indeed, we note that some of the most massive stars in the cluster are present in each of the radial bins.  The binary fraction is more or less uniform, except beyond 30~pc where it drops by about a factor of two.  The velocities of the ejected stars in the halo increase roughly linearly with distance beyond 3~pc (i.e. faster moving stars have travelled further).}
\end{table*}

\begin{figure*}
\includegraphics[width=160mm]{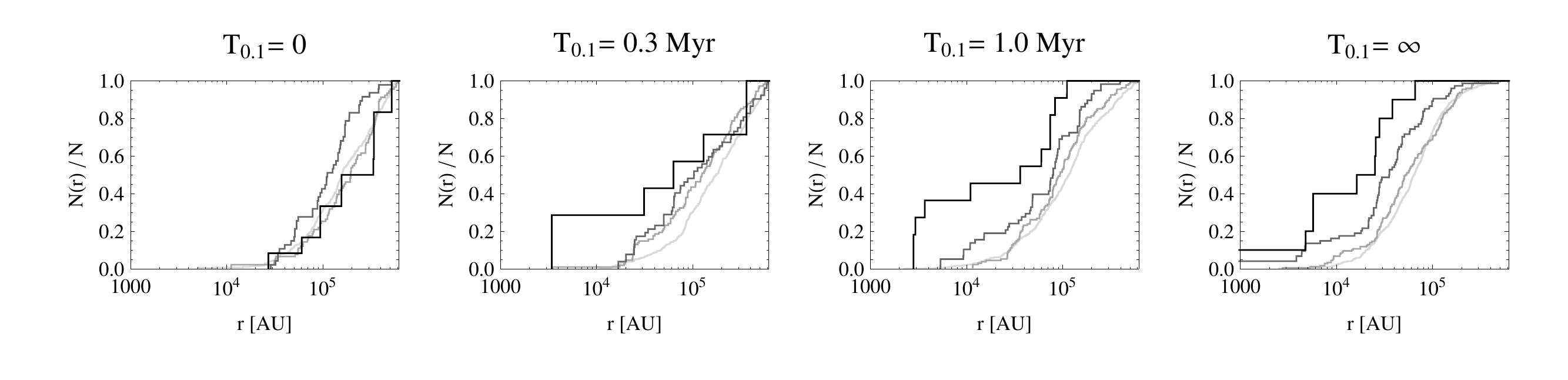}
 \caption{The dependence of the mass segregation at 10~Myr on the gas removal timescale.  From left to right, the four calculations are: instantaneous gas removal, $T_{0.1}=0.3$~Myr, $T_{0.1}=1$~Myr, and no gas removal.  The segregation is measured by the cumulative radial distribution of stars in different mass bins, measured from the position of the star in the region of highest stellar density.  The lines, from light grey to black, show stars in the mass bins $M/$\msun$< 0.1$, $0.1 \le M/$\msun$< 0.3$, $0.3 \le M/$\msun$<1.0$, $1.0 \le M/$\msun. The slow and no gas removal cases display classical mass segregation with more massive stars being progressively more centrally condensed (particularly in the two highest mass bins).}
 \label{MassSegPlummer}
\end{figure*}

In Figure \ref{MassSegregation}, we show the state of the cluster's segregation at three times; the beginning of the {\it N}-body simulation, just after 1 Myr when the most massive stars are about to be ejected from the cluster center, and at around 2 Myr, which is representative of later times.  At the earliest time, the cumulative distribution shows very little evidence of mass segregation \citep[as noted by][]{Bate2009a}, while the surface density analysis shows that the 10 most massive stars have a surface density higher than the 3$\sigma$ value expected for the cluster.  At around 1 Myr, the cumulative distribution shows a clear difference in the radial distribution of stars above 1 \msun.  The surface density of the four most massive stars is still very high, and both plots reflect the lowered surface density of the cluster after its initial expansion.  It is interesting to note that the surface density of the cluster between 1 and 2 Myr has dropped by only a factor of $\sim 2$, while the inner Lagrangian radii (Figure \ref{LagrangianRadii}) have increased by over an order of magnitude, emphasizing that the most massive stars were dominating the inner Lagrangian radii at early times. 
 
By 3 Myr, mass segregation among the four most massive stars remaining in the cluster has reestablished itself in the surface density diagnostic, and remains for the remainder of the simulation.  However, in contrast to the segregation at 1 Myr, this is due to the presence of two massive and well-separated binaries, and not a real clustering of four stars.  The disappearance and return of the mass segregation signal is dependent on what one considers part of the cluster versus the field.  In this case, when dispersing massive stars cease to be considered part of the cluster (by our 3 pc selection criterion), the remaining massive stars are left in a configuration that appear to be clustered.  In addition to the cluster membership determination, the amount of segregation seen in a cluster depends on the viewing angle.  A projection that minimizes the separation between two massive binaries will overemphasize the clustering.  This is important to bear in mind when considering clustering involving only a handful of massive stars.

In Table \ref{tablecluster}, we examine the radial properties of the remnant cluster and halo of ejected stars using logarithmically-spaced bins.  In agreement with the diagnostics discussed above, there is no evidence of mass segregation.  Some of the most massive stars in the system are found in each radial bin.  The velocities of the ejected stars in the halo increase with radius outside of  3~pc from the cluster centre (i.e. faster moving stars have travelled further from the cluster centre).

All four mass-removal scenarios begin with the same initial stellar positions and, therefore, all begin with the 10 most massive stars being centrally concentrated.  While this primordial segregation is transient and only to be observed in the earliest stages of cluster evolution, the more compact remnant clusters that result from slower gas removal have a direct effect on the later mass segregation.  Calculated at 10 Myr, the half-mass relaxation times of the calculations with instantaneous gas removal, $T_{0.1}=0.3$~Myr, $T_{0.1}=1$~Myr, and no gas removal are 20.8~Myr, 12.3~Myr, 8.2~Myr, and~3.3 Myr, respectively.  Thus, whereas the main calculation is dynamically unrelaxed, the two calculations with slow or no gas removal have evolved for more than one relaxation time, which is the timescale on which mass segregation manifests itself.  This is the classical form of mass segregation, due to two-body effects in an almost spherical cluster, in contrast to the primordial segregation that is an imprint of star formation processes and clustered star formation. As such it should appear in cumulative distribution plots as a general trend, with more massive populations progressively more centrally concentrated.

In Figure \ref{MassSegPlummer}, we show the cumulative radial distributions for the main calculation and the three cases with gas potentials at 10~Myr.  The case with $T_{0.1} = 0.3$~Myr shows the expected signal, with the distribution of the most massive stars beginning to be separated from the lower mass stars.  The calculations with $T_{0.1} = 1$~Myr and without gas removal show segregation of the second-most massive bin as well.  We stress that while the earliest (earlier than 1~Myr for these clusters) observations of segregation are a result of a forming cluster's morphology, this later type of mass segregation is expected due to the dynamical maturity of these clusters, and does not contain information about the star formation process.

\subsection{The evolution of multiple systems}
\label{multiplicitysection}

\cite{Bate2009a} quantified the fraction of stars and brown dwarfs that were in multiple systems using the multiplicity fraction, defined as
\begin{equation}
MF = \frac{B+T+Q}{S+B+T+Q},
\end{equation}
where $S$ is the number of single stars within a given mass range and, $B$, $T$, and $Q$ are the numbers of binary, triple, and quadruple systems, respectively, for which the primary has a mass in the same mass range.  This differs from the companion star fraction, $csf$, that is also often used and where the numerator has the form $B+2T+3Q$.  \citeauthor{Bate2009a} chose the multiplicity fraction following \citet{HubWhi2005} who pointed out that this measure is more robust observationally in the sense that if a new member of a multiple system is found (e.g. a binary is found to be a triple) the quantity remains unchanged.  It is also more robust for simulations too in the sense that if a high-order system decays because it is unstable the numerator only changes if a quadruple decays into two binaries (which is quite rare).  Furthermore, if the denominator is much larger than the numerator (e.g. for brown dwarfs where the multiplicity fraction is low) the production of a few single objects does not result in a large change to the value of $MF$.  This was useful for \cite{Bate2009a} because many of the high-order systems in existence at the end of the calculations were likely to undergo further dynamical evolution.  Indeed, the long-term evolution of the multiple systems produced by the main calculation of \cite{Bate2009a} is the topic of this section.

\begin{figure*}
\centering
    \includegraphics[width=8.4cm]{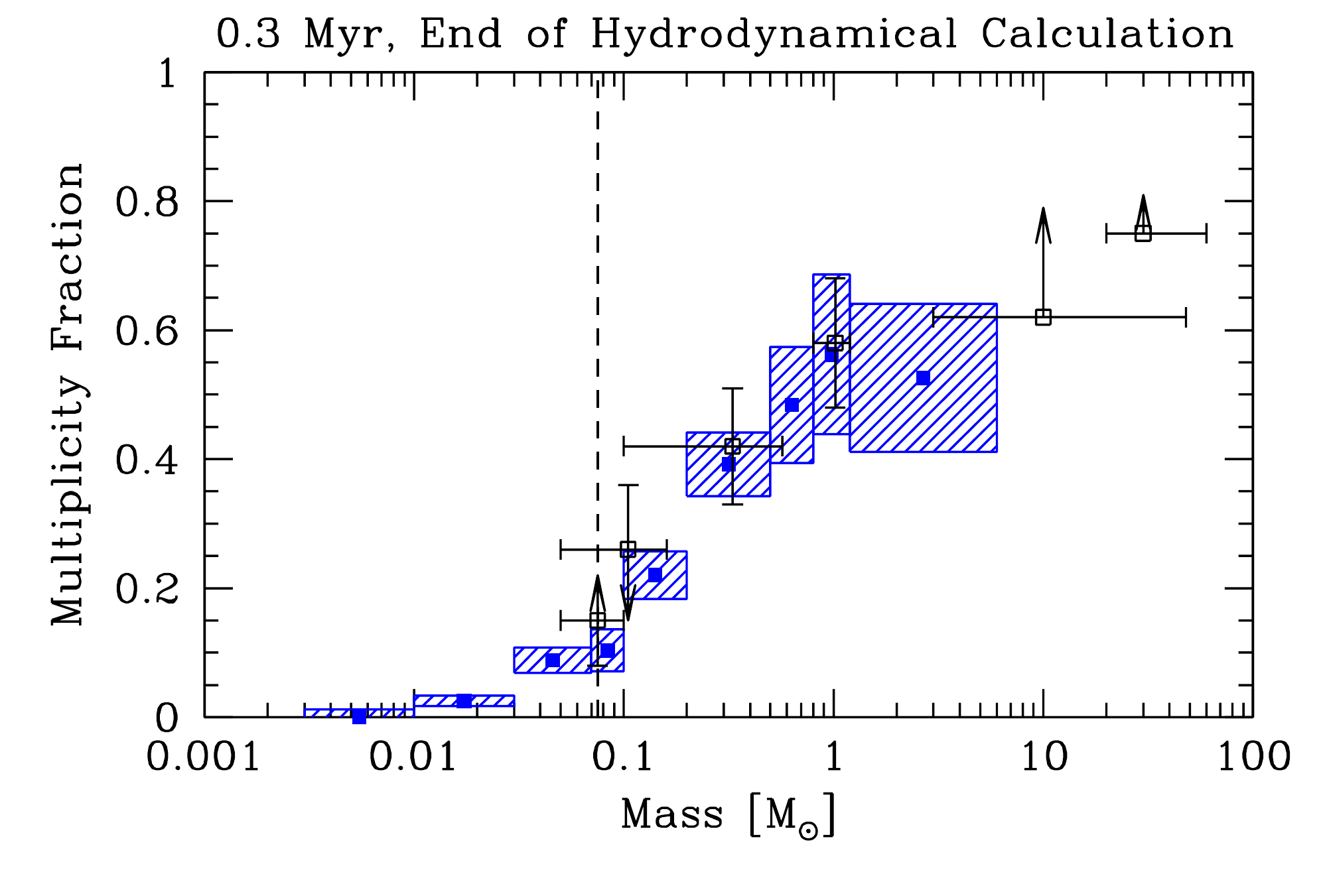}
    \includegraphics[width=8.4cm]{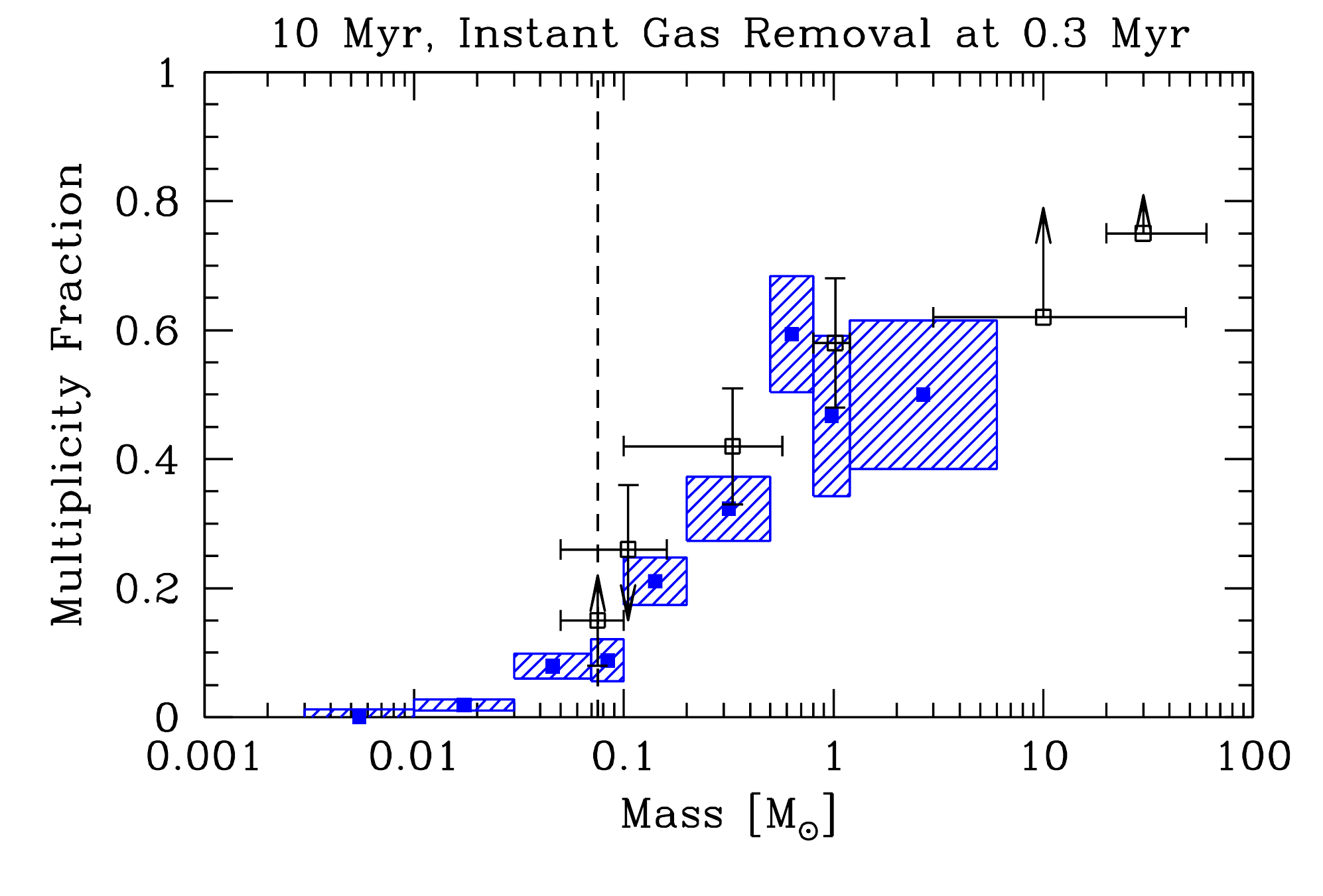} \vspace{0cm}
    \includegraphics[width=8.4cm]{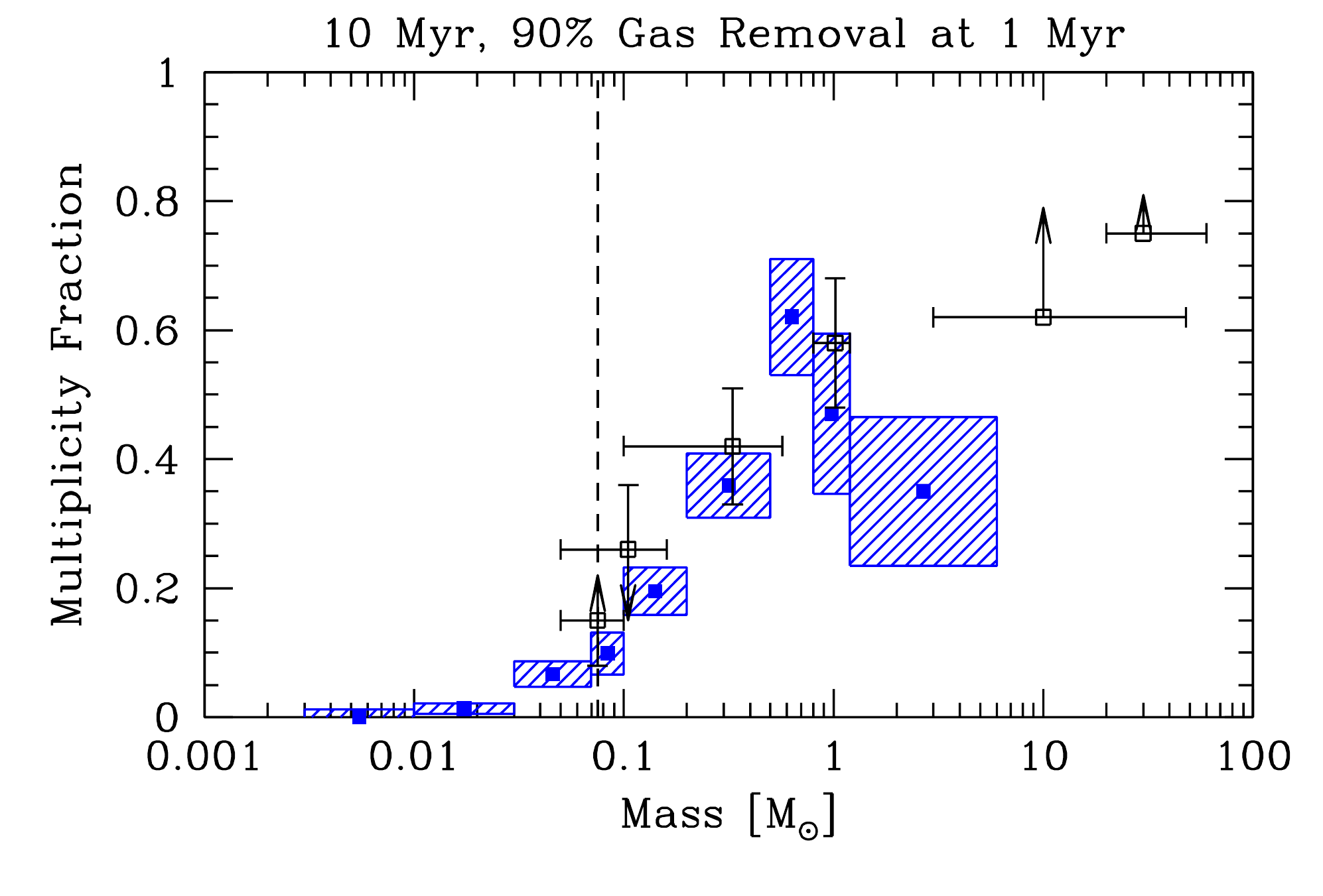}
    \includegraphics[width=8.4cm]{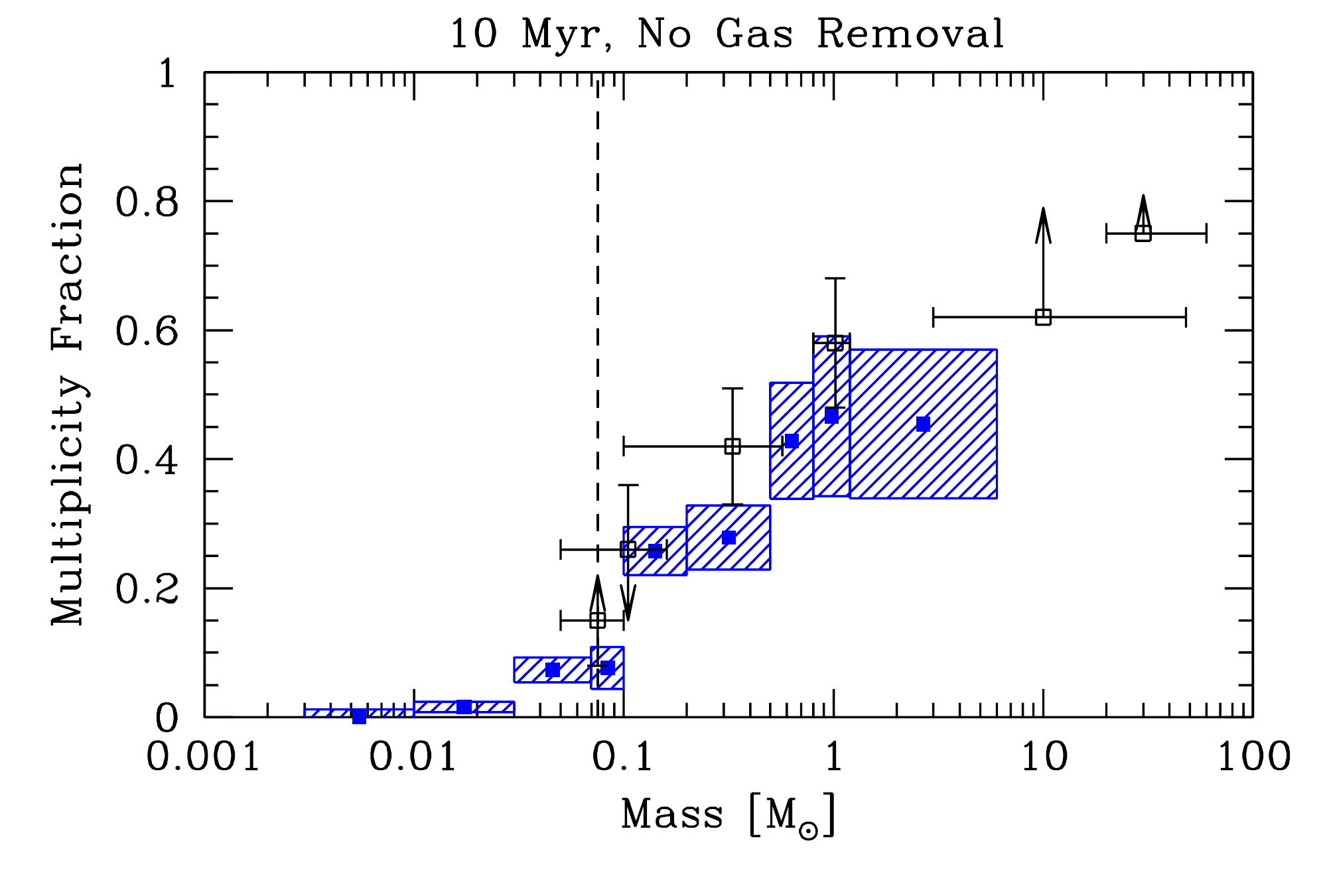} \vspace{0cm}
\caption{Multiplicity fraction as a function of primary mass.  The top left panel gives the results at the end of the hydrodynamical calculation (0.3~Myr), while the other panels give the results at the end of the {\it N}-body evolutions (10~Myr) for three cases: instantaneous gas removal (top right), $T_{0.1}=0.3$~Myr (lower left), and with no gas removal (lower right).  The blue filled squares surrounded by shaded regions give the results from the calculations with their statistical uncertainties.  The open black squares with error bars and/or upper/lower limits give the observed multiplicity fractions from the surveys of Close et al. (2003), Basri \& Reiners (2006), Fisher \& Marcy (1992), Duquennoy \& Mayor (1991), Preibisch et al. (1999) and Mason et al. (1998), from left to right.  There is very little evolution of the multiplicity fraction over the 10~Myr of {\it N}-body evolution, even for the extreme case of no gas removal.  All sets of results are in reasonable agreement with the observed multiplicity as a function of primary mass. }
\label{multiplicities}
\end{figure*}

In Figure \ref{multiplicities}, we plot the multiplicity fraction of the stars and brown dwarfs as a function of stellar mass for the calculation at the end of the hydrodynamical calculation (i.e. the beginning of the {\it N}-body evolution) and at the end of the main {\it N}-body calculation at an age of 10 Myr (top left and top right panels, respectively).  The top left panel is a reproduction of Figure 15 of \cite{Bate2009a}.  The mass bins were chosen for comparison with the results of various observational surveys.  The filled blue squares give the multiplicity fractions from the calculation while the surrounding blue hatched regions give the range in stellar masses over which the fraction is calculated and the $1\sigma$ (68\%) uncertainty on the multiplicity fraction.  The black open boxes and their associated error bars and/or upper/lower limits give the results from a variety of observational surveys (see the figure caption).

There is very little evolution of the multiplicity fraction during the main {\it N}-body evolution of the cluster with instantaneous gas removal.  Both the results at the end of the hydrodynamical calculation and those at the end of the {\it N}-body evolution are in good quantitative and qualitative agreement with the observed multiplicity fraction (see \cite{Bate2009a} for a full discussion), which is found to be a strongly increasing function of primary mass.  We also note that the decrease in binary fraction in the outer parts of the ejected halo of stars found by \cite{Bate2009a} is maintained to ages of 10~Myr (Table \ref{tablecluster}), but that the transition radius at which the binary fraction begins to decrease with distance has moved from 0.15~pc to 30~pc as the halo has expanded.

For the randomly-perturbed sets of initial conditions, although each calculation differs in detail with respect to which binaries and multiple systems survive or evolve dynamically, any variations in the values of the multiplicity as a function of primary mass are smaller than the statistical uncertainties already plotted in Figure \ref{multiplicities}.  Typically, for each mass range one or two more or fewer multiple systems survives in each random realisation for primary masses $<0.50$~M$_\odot$.  For the more massive systems there tends to be somewhat more variation.  For example, for the most massive systems (primary masses $>1.2$~M$_\odot$) the multiplicity ranges from $6/22 = 27$\% to $10/20=50$\%.  But the statistical uncertainties are also largest for these multiplicities due to the small numbers of massive stars.

The lower panels of Figure \ref{multiplicities} give the multiplicity fractions for the two of the calculations aimed at investigating the dependence of the results on the gas removal time.  The results from the $T_{0.1}=1$~Myr (lower left panel) and no gas removal (lower right panel) calculations are given.  It can be seen that the multiplicity fractions decrease slightly overall with increased gas removal timescales, but only marginally despite the fact that the remnant clusters become significantly more condensed and populous.  Even in the extreme case of no gas removal, the decreases in the multiplicity fractions for different primary masses between the instantaneous gas removal and no gas removal cases are smaller than the statistical uncertainties.  As for the randomly-perturbed sets of initial conditions, only the most massive systems (primary masses $>1.2$~M$_\odot$) have a large change in the multiplicity fraction, but, again, the statistical uncertainties are also largest in this mass range due to the small numbers of massive stars.

\subsubsection{Triple and quadruple systems}

Although the multiplicity fraction does not evolve significantly, this is partially due to the fact that this measurement of the fraction of binaries and multiples was deliberately chosen to be relatively insensitive to the break up of dynamically-unstable multiple systems.  In fact, during the {\it N}-body evolution of the unperturbed case with instantaneous gas removal, the number of quadruple systems drops from 25 to 8, while the number of triple systems remains the same (though the specific make up of the 23 systems changes).  The number of binary systems increases slightly from 90 to 98 and the number of single objects also increases from 904 to 956 (Table \ref{summarytable}).  Much of this evolution occurs quite quickly, with the number of quadruple systems having dropped to 10 after 0.3~Myr of {\it N}-body evolution (i.e. when the cluster age is 0.6~Myr).  Thus, high-order systems (particularly quadruple systems) are converted into lower-order systems through dynamical disruption, but the multiplicity fractions themselves are not significantly changed.  

The final overall frequencies of triple and quadruple systems are 2.1 and 0.7\%, respectively.  However, as discussed by \cite{Bate2009a}, these frequencies are a strong function of primary mass.  At the end of the {\it N}-body evolution there are no quadruple systems with primary masses less than 0.2~M$_\odot$.  There are two surviving very-low-mass (VLM) ($0.03-0.10$~M$_\odot$) triple systems out of a total of 331 systems (giving a frequency of 0.6\%).  At the end of the hydrodynamical calculation there were 3 triples and 2 quadruple systems with primaries in this mass range.  Thus, the fraction of high-order VLM multiples has significantly decreased.  For low-mass stars, there are 4 triple systems out of the 133 systems with primaries in the range $0.1-0.2$~M$_\odot$ (i.e. 3\%) at the end of the {\it N}-body evolution, whereas at the end of the hydrodynamical evolution there were 4 triples and 2 quadruples.  The frequencies of triples or quadruples for M-dwarf primaries and masses $0.2-0.5$~M$_\odot$ is 9 out of 96 systems (5 triples and 4 quadruples), very similar to the 7 triples and 1 quadruple found from the 99 M-dwarf primaries surveyed by \cite{FisMar1992}.  At the end of the hydrodynamical evolution there were 5 triples and 10 quadruples with primaries in this mass range.  For higher-mass primaries, the frequencies of triple and quadruple systems are 18\% for $0.5-0.8$~M$_\odot$ and 29\% for primary masses $>0.8$~M$_\odot$, but with large statistical uncertainties.  The number of high-order systems with these higher mass primaries is similar after the {\it N}-body evolution to that at the end of the hydrodynamical evolution, but whereas the number of triples has increased from 10 to 12, the number of quadruples has dropped from 11 to 4.

The destruction of high-order multiple systems is qualitatively the same for the randomly-pertubed realisations and even for the calculations that include a gas potential.  In Table \ref{summarytable}, we give the total numbers of single, binary, triple and quadruple systems at the end of the hydrodynamical calculation and at the end of the four {\it N}-body calculations with different gas removal timescales.  For gas removal timescales of 1~Myr or less, the numbers are statistically indistinguishable.  For the calculation without gas removal, the number of triples is lower, but on the other hand the number of quadruples is somewhat higher than the in the cases with gas removal.  Closer inspection (see the next section) shows that many of these quadruples are very wide systems in the halo.

\begin{figure}
\centering
    \includegraphics[width=4.4cm]{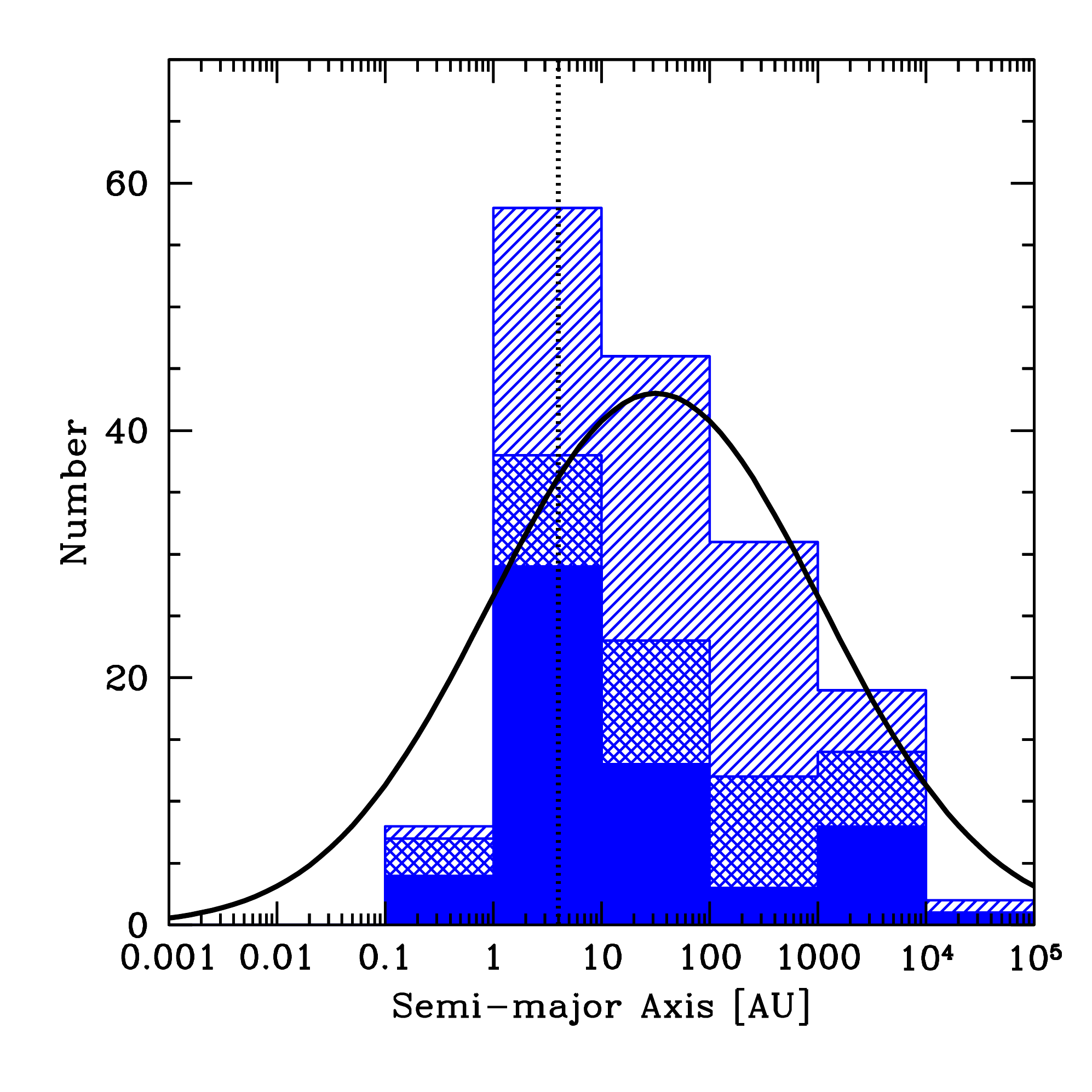} \hspace{-0.5cm}\vspace{-0.2cm}
    \includegraphics[width=4.4cm]{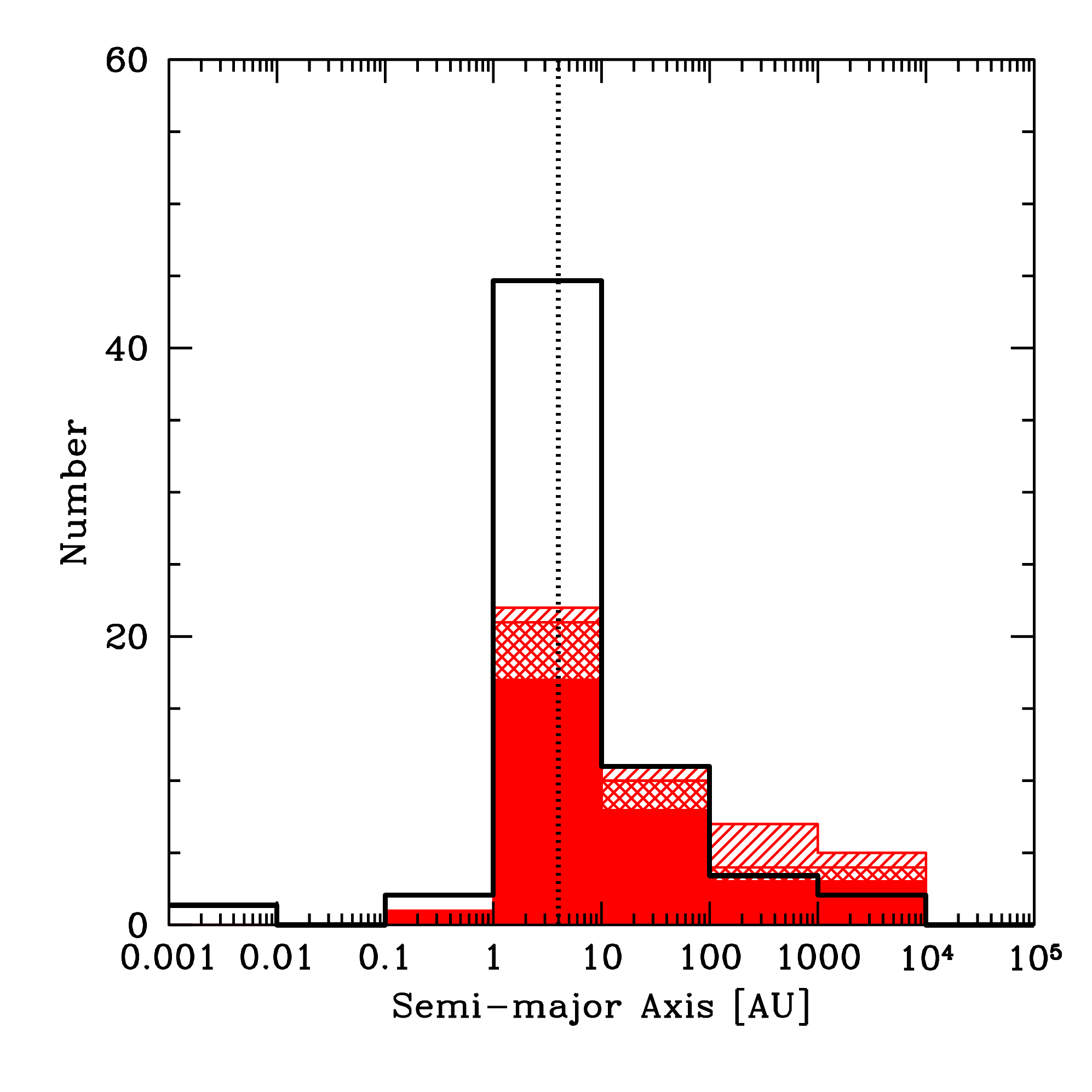} \vspace{-0.2cm}
    \includegraphics[width=4.4cm]{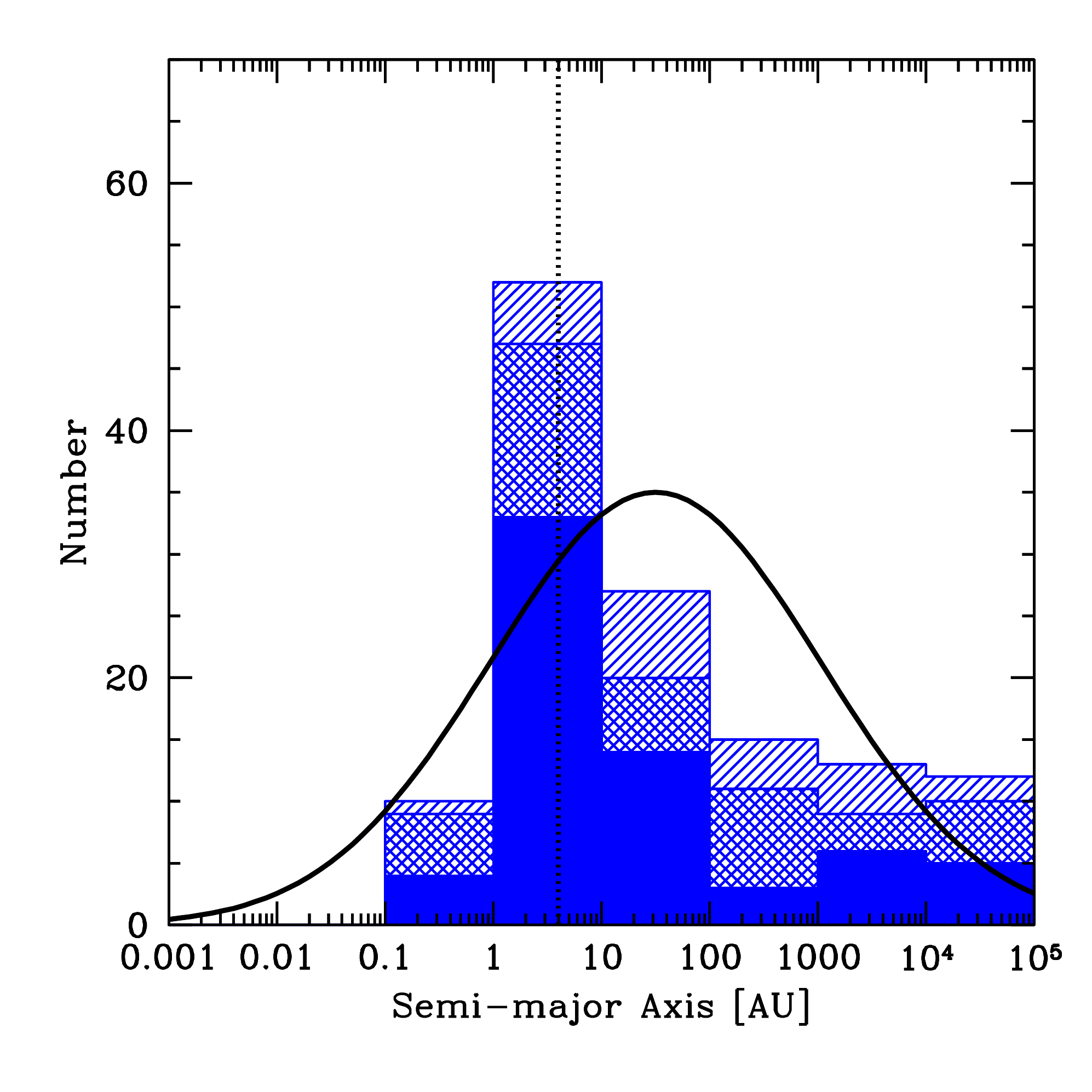} \hspace{-0.5cm} 
    \includegraphics[width=4.4cm]{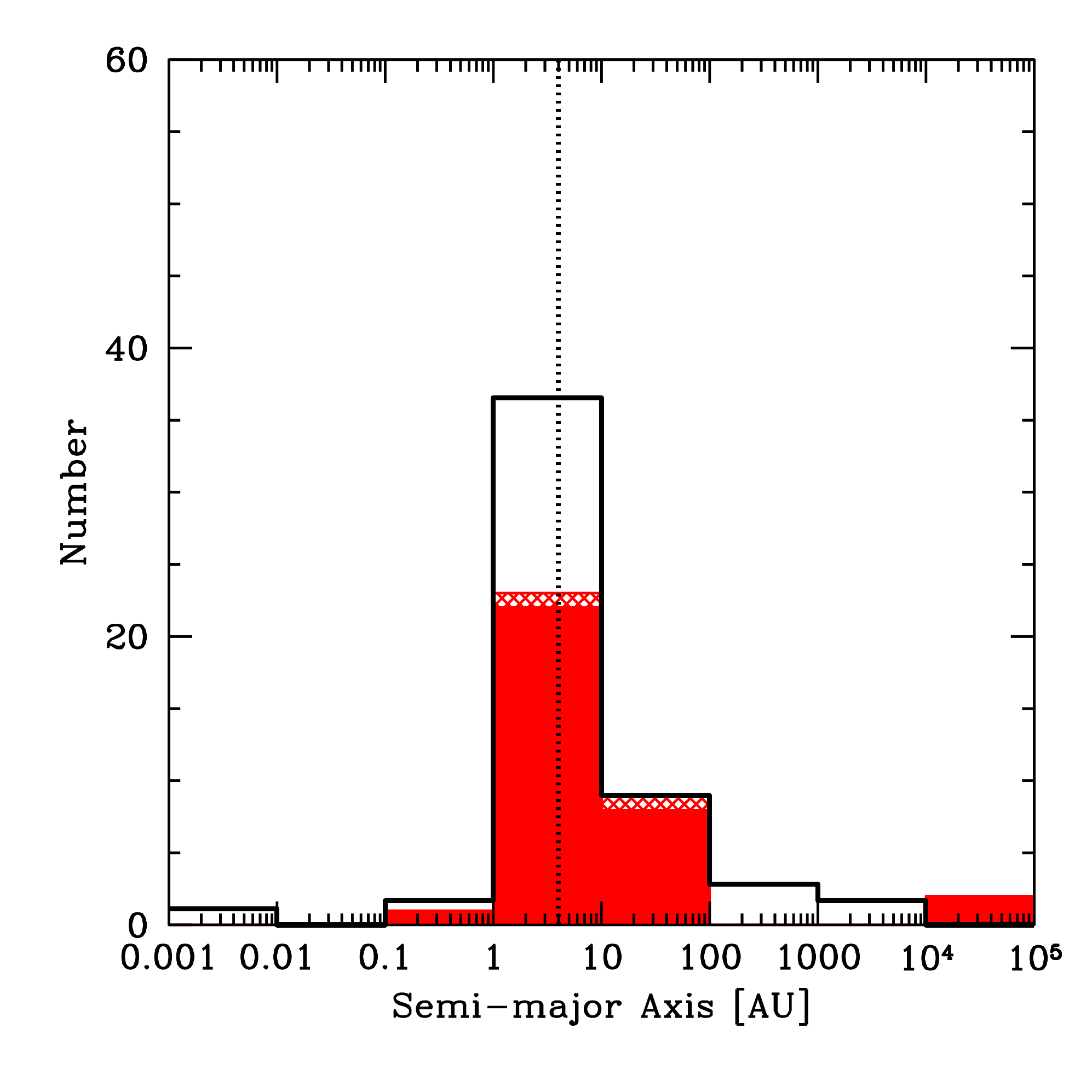}  \vspace{-0.2cm}
    \includegraphics[width=4.4cm]{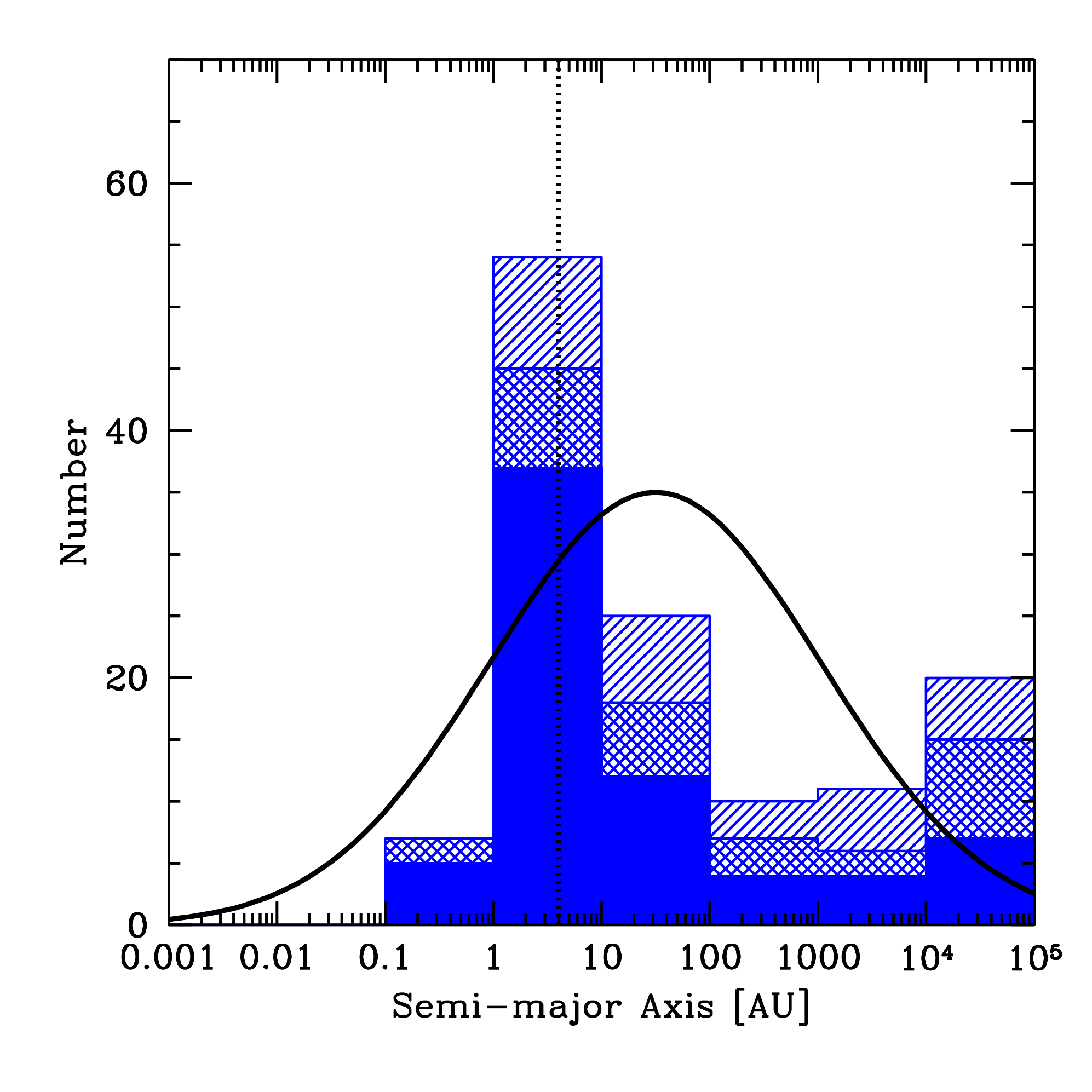} \hspace{-0.5cm} 
    \includegraphics[width=4.4cm]{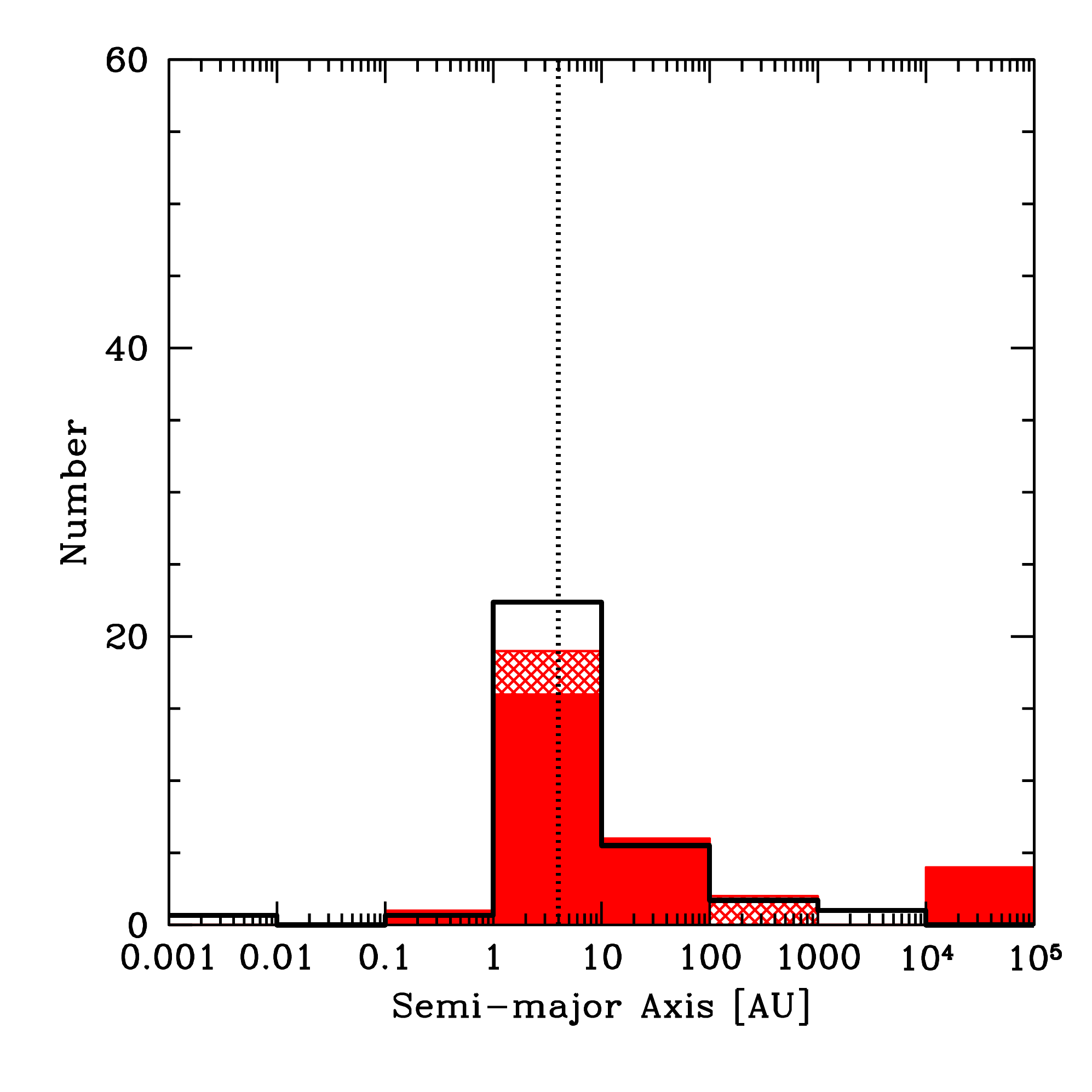}  \vspace{-0.2cm}
    \includegraphics[width=4.4cm]{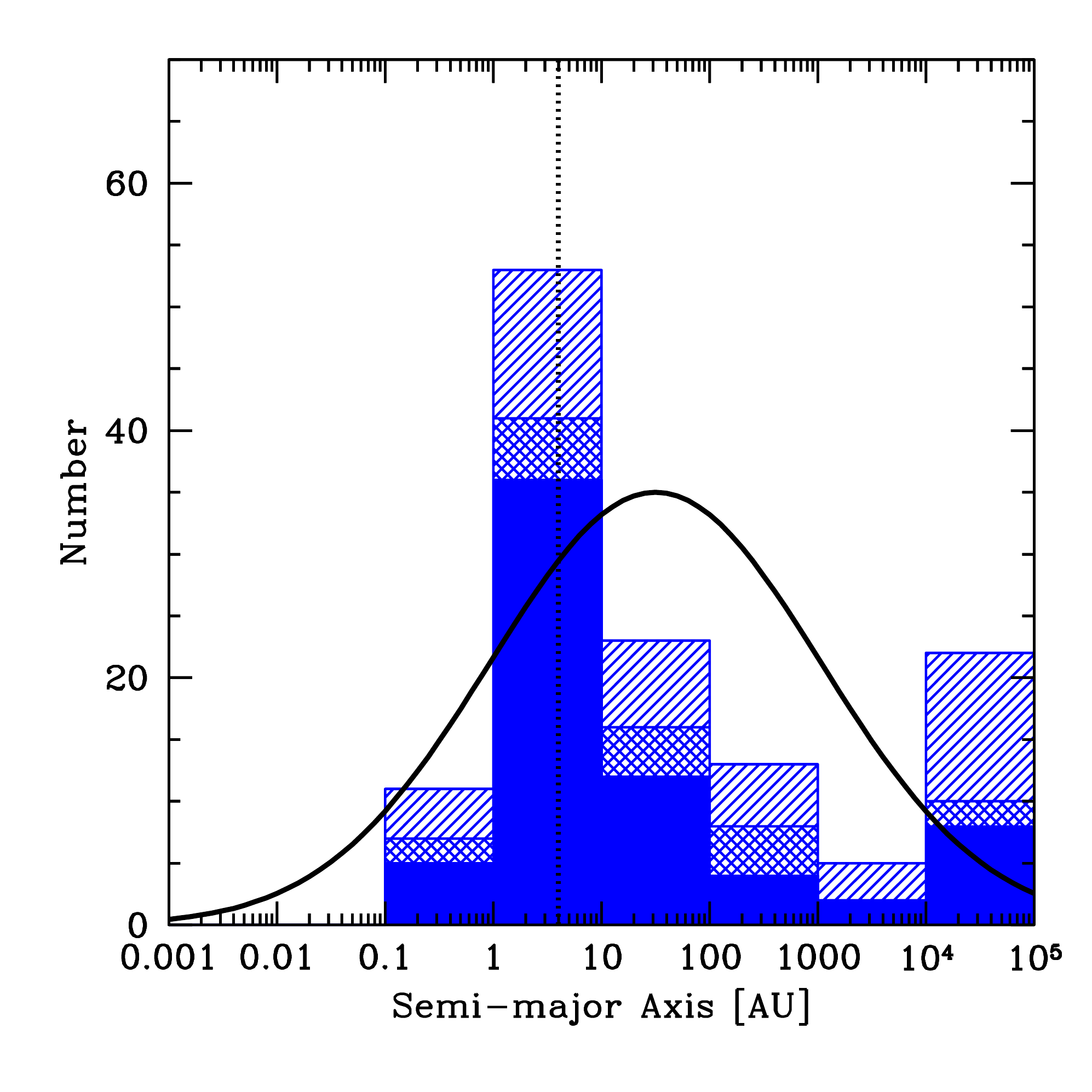} \hspace{-0.5cm} 
    \includegraphics[width=4.4cm]{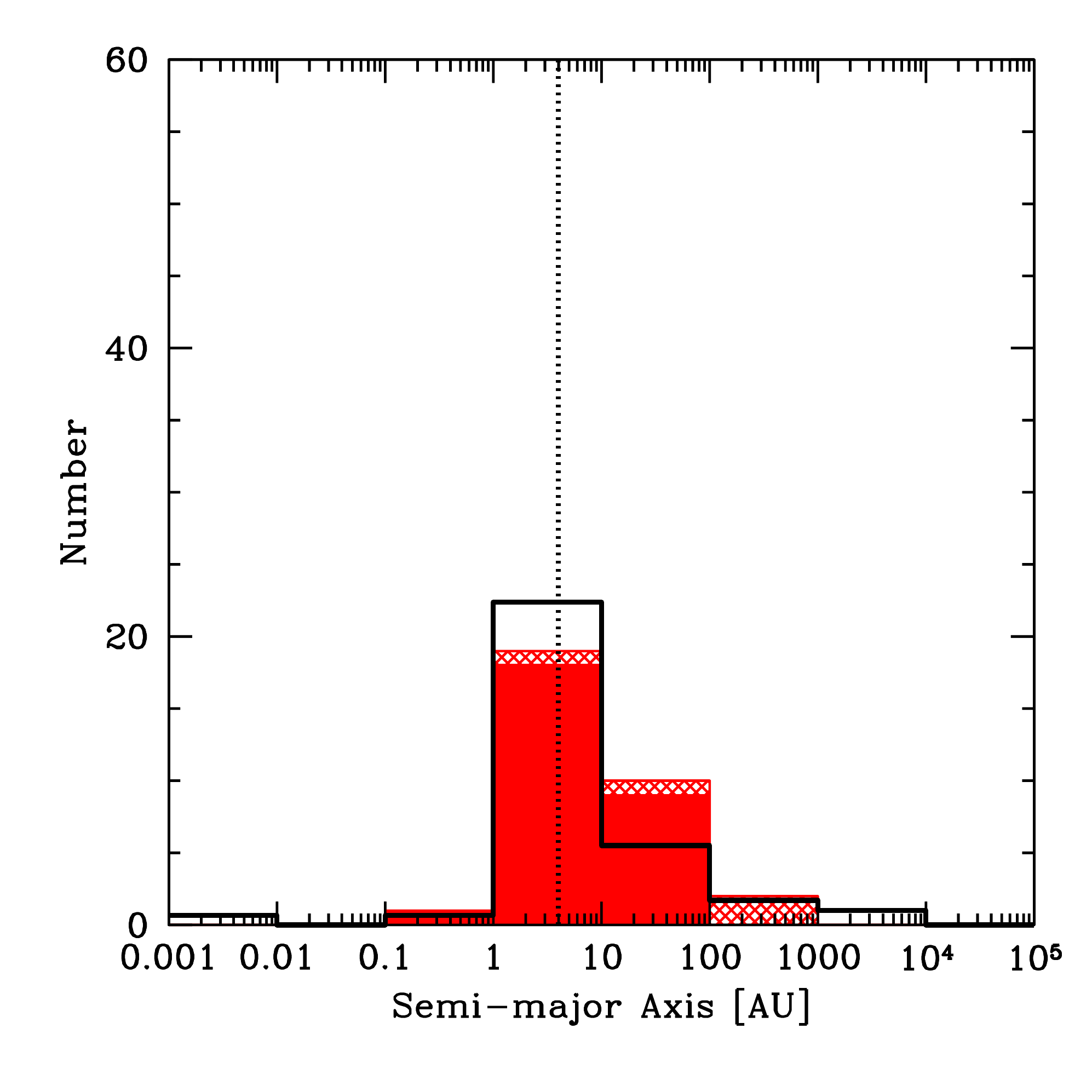} 
\caption{The distributions of separations (semi-major axes) of multiple systems with stellar (left) and VLM (right) primaries.  From top to bottom, the four rows give the results: at the end of the hydrodynamical calculation (top row; 0.3~Myr; reproduced from Bate 2009a) and at the end of the {\it N}-body evolution (10~Myr) with instantaneous gas removal (second row), gas removal timescale $T_{0.1}=1$~Myr (third row), and no gas removal (bottom row). The $T_{0.1}=0.3$~Myr case has been omitted as the results lie between those of the instantaneous gas removal and $T_{0.1}=1$~Myr cases.  The solid, double hashed, and single hashed histograms give the orbital separations of binaries, triples, and quadruples, respectively (each triple contributes two separations, each quadruple contributes three separations).  In the stellar graph, the curve gives the G-dwarf separation distribution (scaled to match the area under the histogram in each case) from Duquennoy \& Mayor (1991).  In the VLM systems graph, the open black histogram gives the separation distribution (scaled to match the number in the 10-100 AU range of the calculated histogram in each case) of the known VLM multiple systems maintained by N. Siegler at http://vlmbinaries.org/ (last updated on February 4th, 2008).  The vertical dotted line gives the resolution limit of the hydrodynamical calculation as determined by the gravitational softening and accretion radii of the sink particles.}
\label{separations}
\end{figure}

\subsubsection{Separation distributions}

Along with the evolution of the frequencies of multiple systems, we wish to understand how the characteristics of the multiple systems evolve.  In Figure \ref{separations}, we present the separation (semi-major axis) distributions of the stellar (primary masses greater than 0.10 M$_\odot$; left panels) and VLM multiples (right panels).  These distributions are compared with the survey of \citet{DuqMay1991} and the listing of VLM multiples maintained by N.\ Siegler, C.~Gelino, and A.~Burgasser at http://vlmbinaries.org/ (last updated July 28, 2009), respectively.  The filled histograms give the separations of binary systems, while the double hashed region adds the separations from triple systems (two separations for each triple, determined by sub-dividing the triple into a binary with a wider companion), and the single hashed region includes the separations of quadruple systems (three separations for each quadruple which may be composed of two binary components or a triple with a wider companion).  The top row gives the distributions at the end of the hydrodynamical calculation, while the second row gives the distributions at the end of the {\it N}-body calculation (10~Myr) with instantaneous gas removal.

We note that the gravitational softening and finite sink particle accretion radii used in the hydrodynamical calculation of \cite{Bate2009a} result in a pile up of stellar binaries with separations of a few AU.  This was investigated by \citeauthor{Bate2009a} who re-ran part of the hydrodynamical calculation without gravitational softening and using smaller accretion radii and, as expected, the pile up disappeared and a bell-shaped separation distribution was recovered (but with a much smaller number of objects because the re-run calculation could not be followed for as long).

Comparing the upper two left-hand panels in Figure \ref{separations} for the stellar multiple systems, we find that, as mentioned earlier, many of the quadruple systems are broken up, resulting in a marked decrease in the contribution of the quadruple systems to the separation distributions.  There is also an increase in the number of triple systems with component separations $<10$~AU.  The other significant change is in the number of very wide systems (separations $10^4-10^5$~AU).  While at the end of the hydrodynamical calculation there were only two separations in this range (one binary, and one from the wide component of a quadruple), at 10~Myr there are 7 binaries, 5 triples, and 2 quadruples with separations in this range.  These wide systems are found in the halo of ejected objects and are formed when two ejected objects have similar velocities and as they depart from the cluster, they happen to be weakly mutually bound. The overall effect of the break up of stellar quadruples with intermediate separations ($1-1000$~AU) and the formation of very wide systems is to significantly broaden the separation distribution.  In fact, the separation distribution from $10-10^5$~AU appears to be even flatter than that observed by \cite{DuqMay1991}, and more consistent with the flat separation distribution of wide ($500-5000$~AU) binaries found for pre-main-sequence stars by \cite{KraHil2009}.

For the VLM systems (primary masses $<0.1$~M$_\odot$), the evolution is less dramatic (upper two right-hand panels of Figure \ref{separations}).  Many of the triples and all of the quadruple systems present at the end of the hydrodynamical calculation have been broken up by the end of the {\it N}-body evolution.  However, these comprise a smaller fraction of the initial separation distribution than for the stellar systems.  There are two main effects of the long-term evolution.  First, all the systems with separations greater than 60 AU at the end of the hydrodynamical calculation have been destroyed.   Second, two extremely wide VLM binaries have been formed in the dispersing halo population: a 26,000 AU system consisting of 40 and 14 M$_{\rm J}$ brown dwarfs and a 34,000 AU system consisting of 39 and 10 M$_{\rm J}$ brown dwarfs.
This is good agreement with the observed separation distribution of VLM binaries: field VLM binaries are almost always found to have projected separations $\lsim 40$~AU \citep{Martinetal2000, Reidetal2001,Closeetal2003,Burgasseretal2007} with only a handful of wider field systems found to date (220 AU; 1800 AU, 5100 AU, 6700 AU; \citealt{Billeresetal2005,Caballero2007, Artigauetal2007, Radiganetal2009}), while 4 out of 5 VLM binaries with separations in the $100-1000$ AU range have been found in star-forming regions \citep{Luhman2004a,JayIva2006,Closeetal2007,Bejaretal2008}. 

The evolution of the separation distributions discussed above is qualitatively similar for all the randomly-perturbed realisations, though quantitatively there is some variation.  The number of wide ($10^4-10^5$ AU) systems displays significant variation.  In the unperturbed case, 7 binary, 5 triple and 2 quadruple separations fell within this separation range.  The mean values for all 11 realisations are: $6.8 \pm 3.5$, $3.6 \pm 2.7$, and $4.5 \pm 2.5$ with the total number of separations in this range being $15 \pm 7.2$ with the minimum being 5 and the maximum being 32.  Thus, the unperturbed case is similar to the mean case, but sometimes there can be a lot more or a lot fewer wide multiple systems formed.  This is not unexpected, since these systems require stars or systems to be ejected with similar velocities, something that is very sensitive to small perturbations.

\begin{figure*}
\centering
    \includegraphics[width=5.0cm]{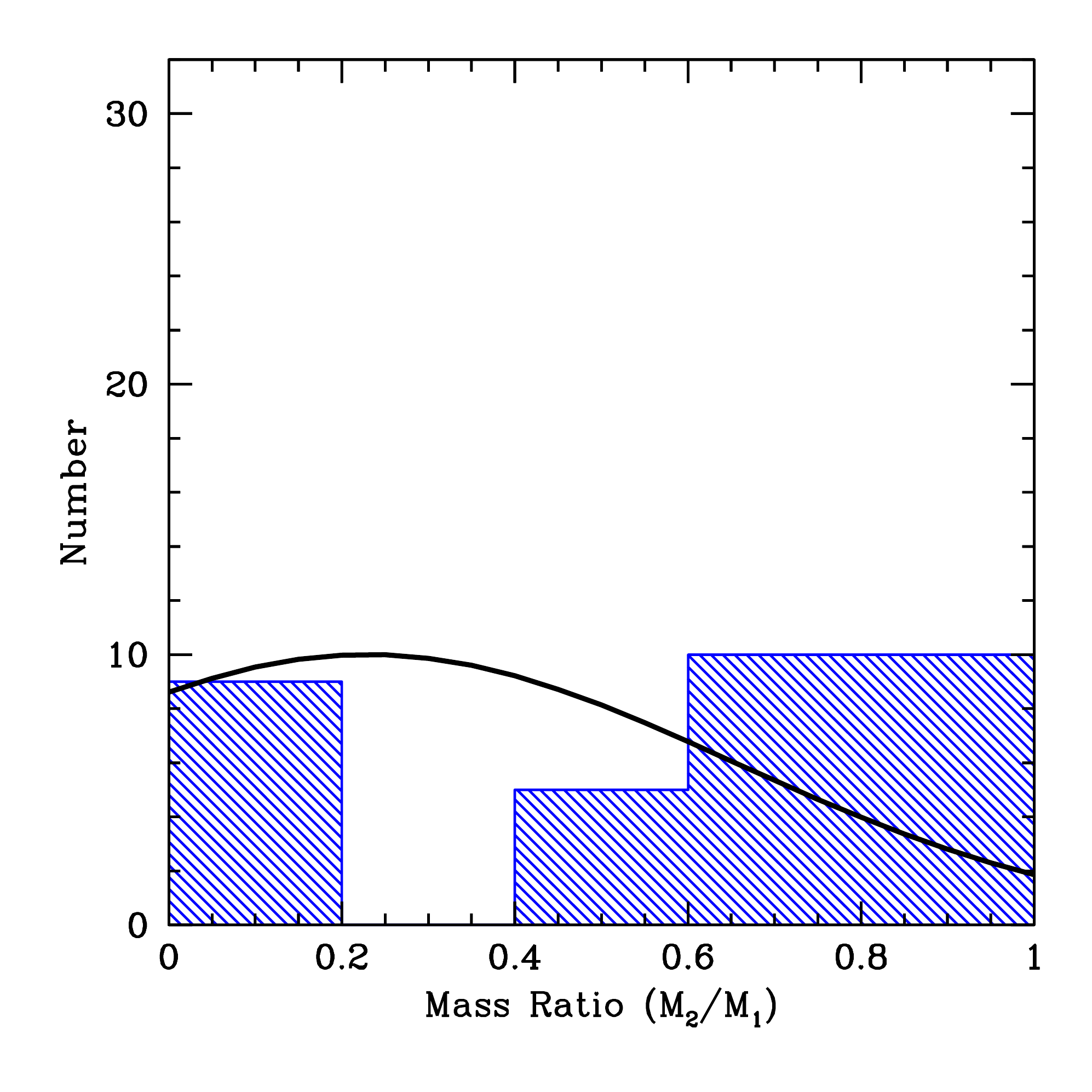}
    \includegraphics[width=5.0cm]{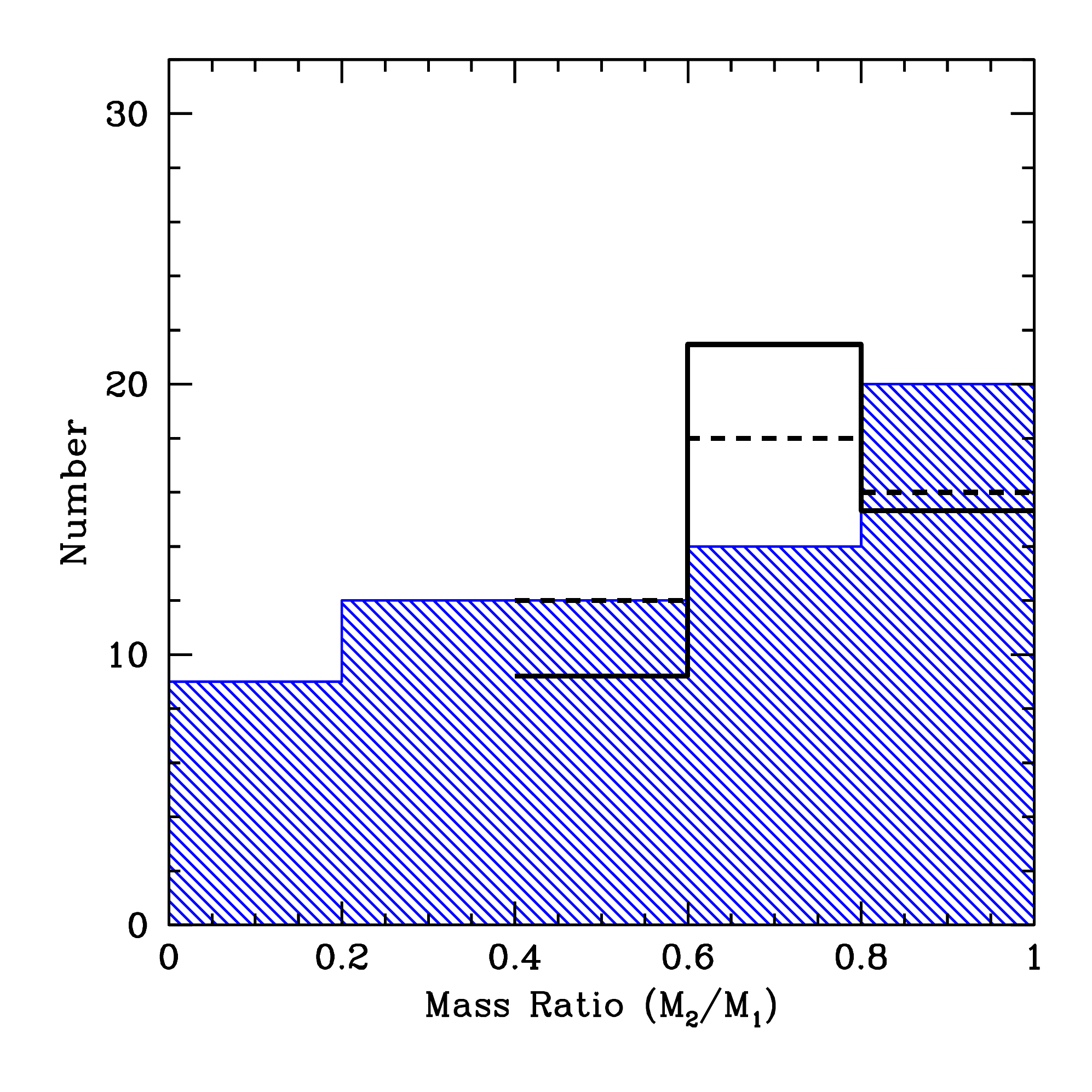}
    \includegraphics[width=5.0cm]{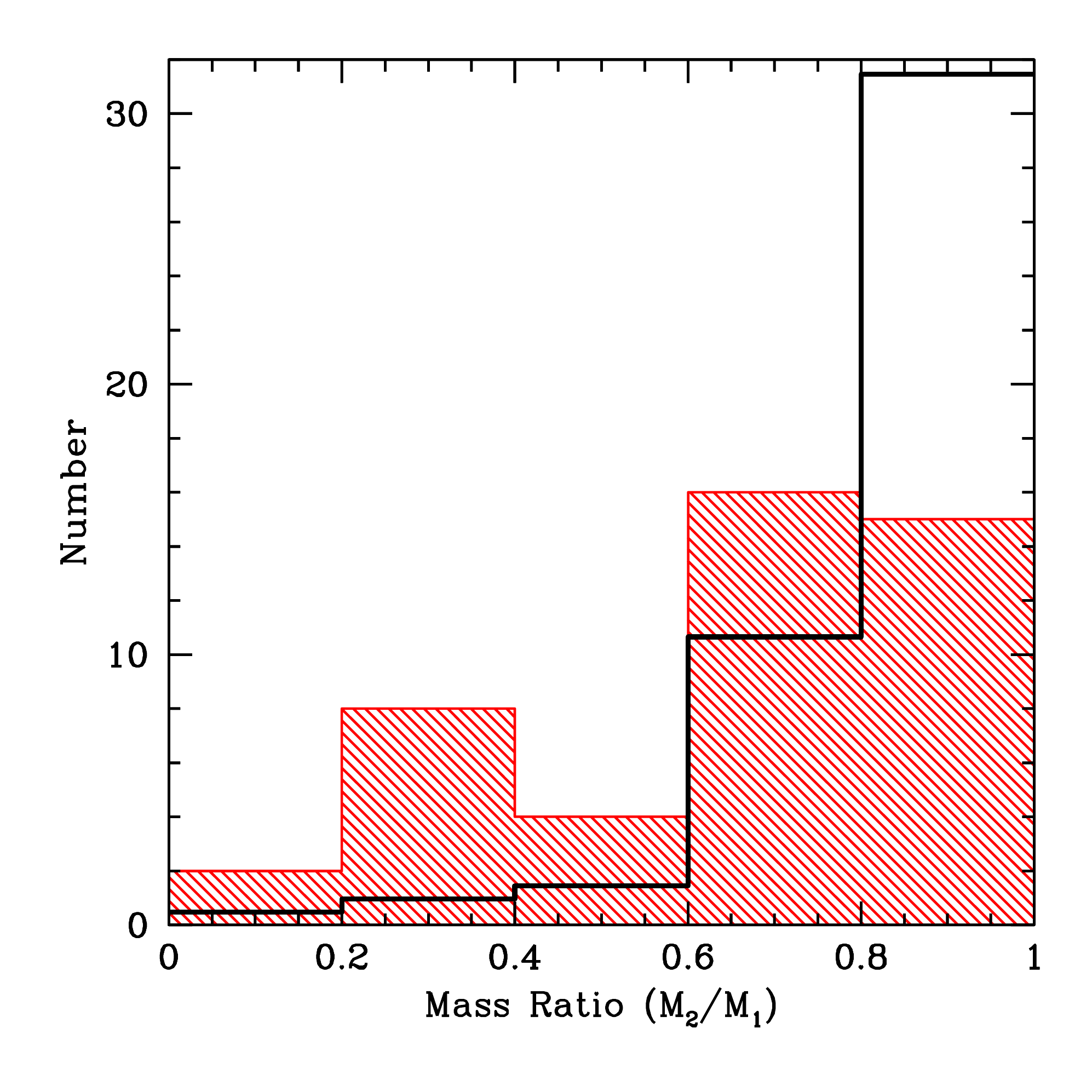}\vspace{-5pt}
    \includegraphics[width=5.0cm]{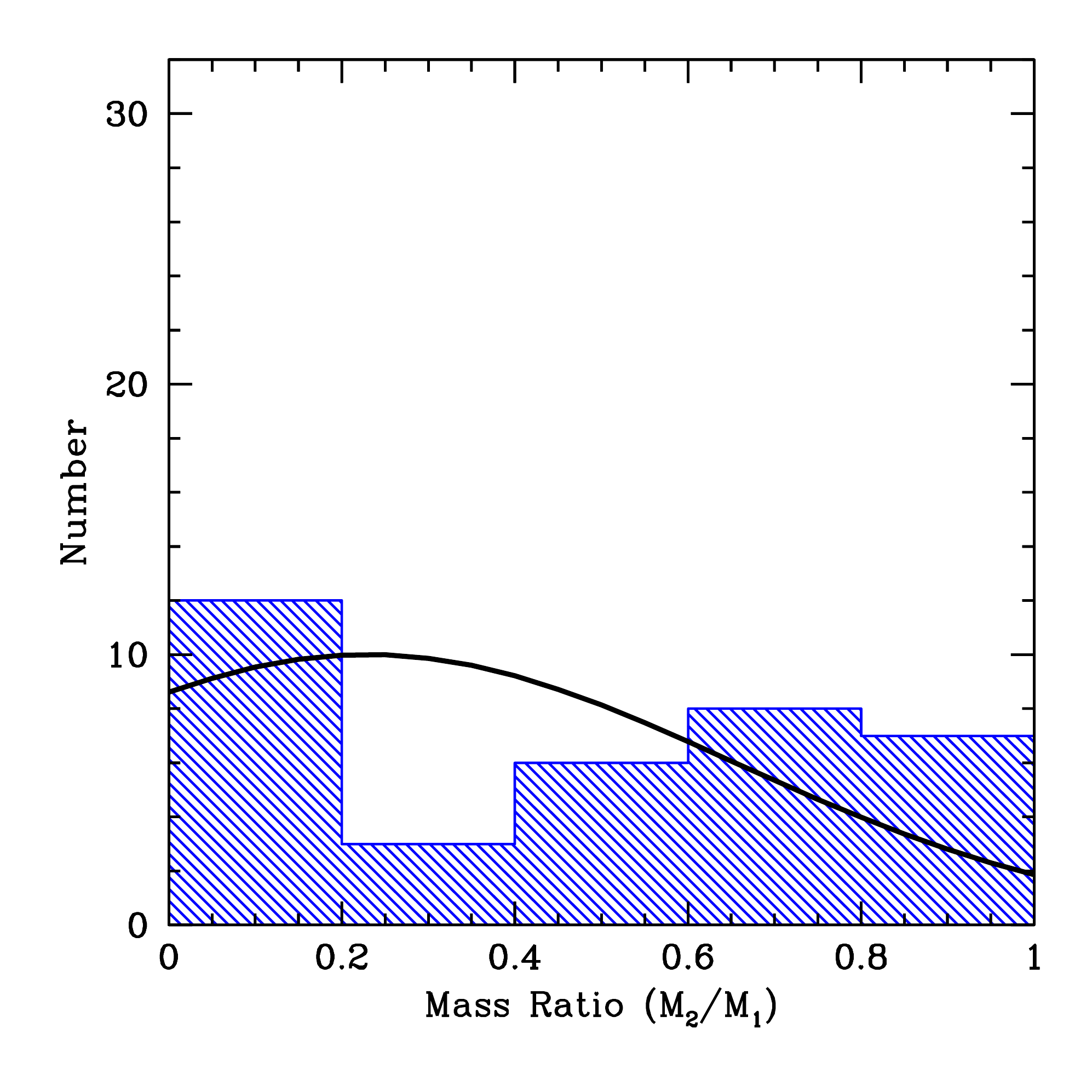}
    \includegraphics[width=5.0cm]{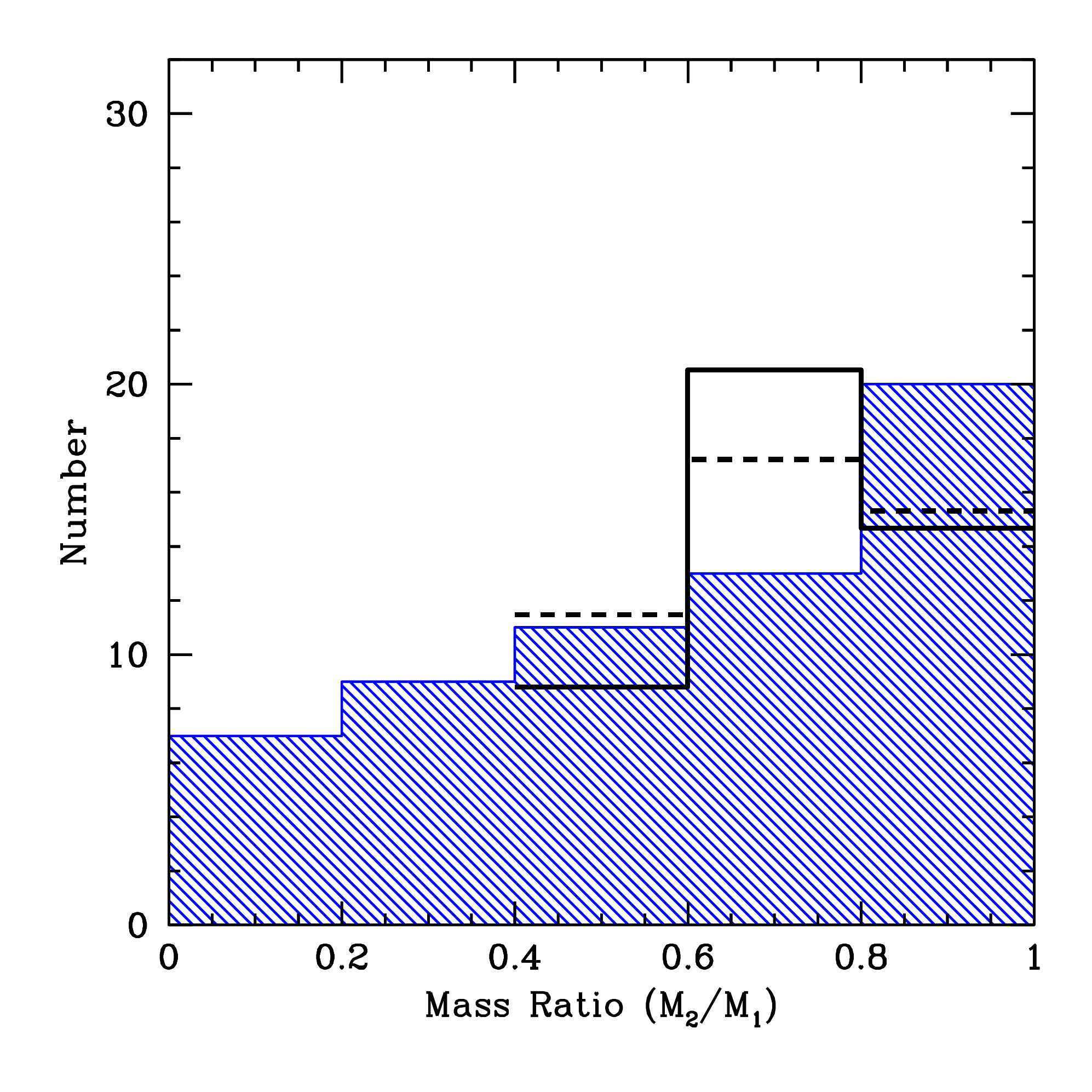}
    \includegraphics[width=5.0cm]{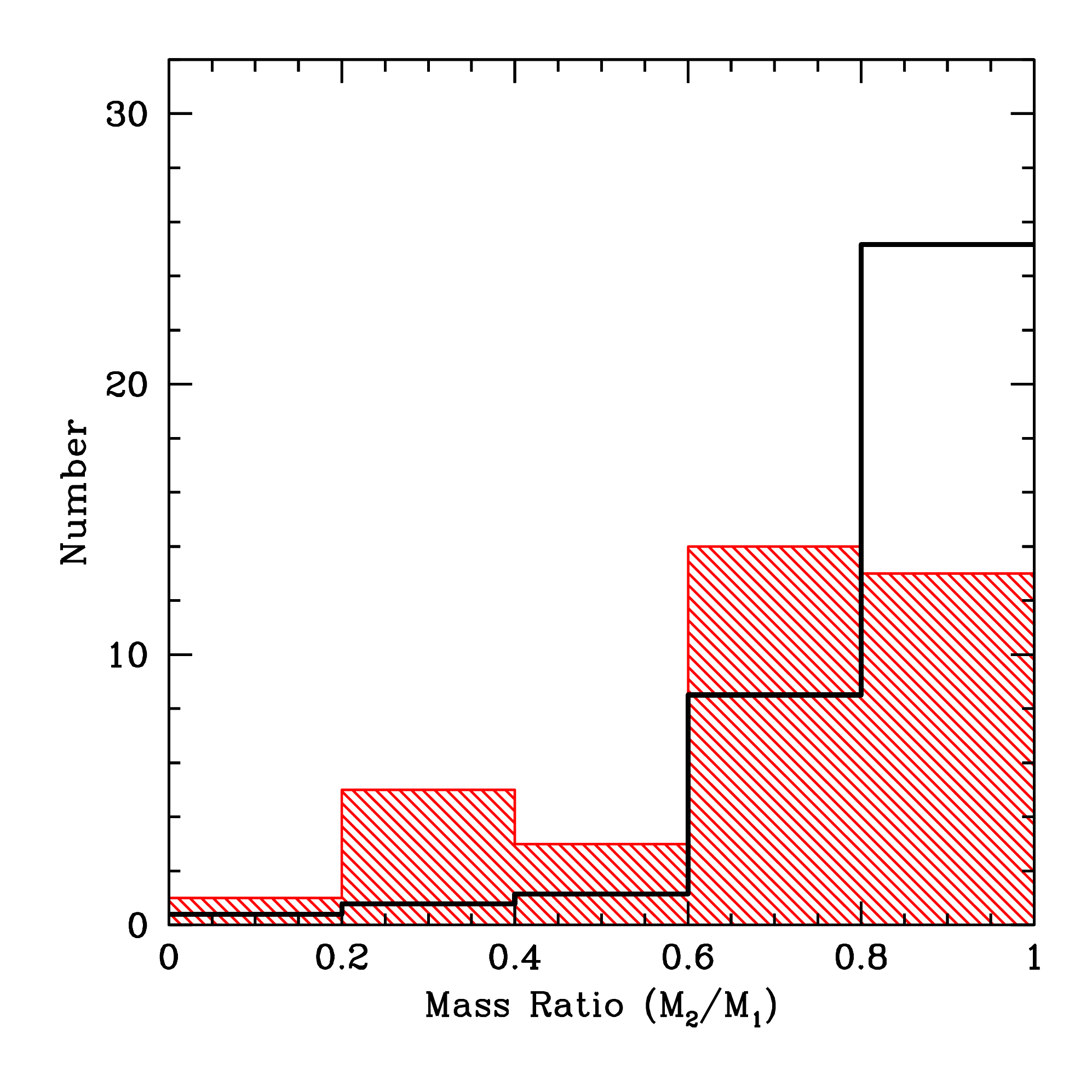}\vspace{-5pt}
    \includegraphics[width=5.0cm]{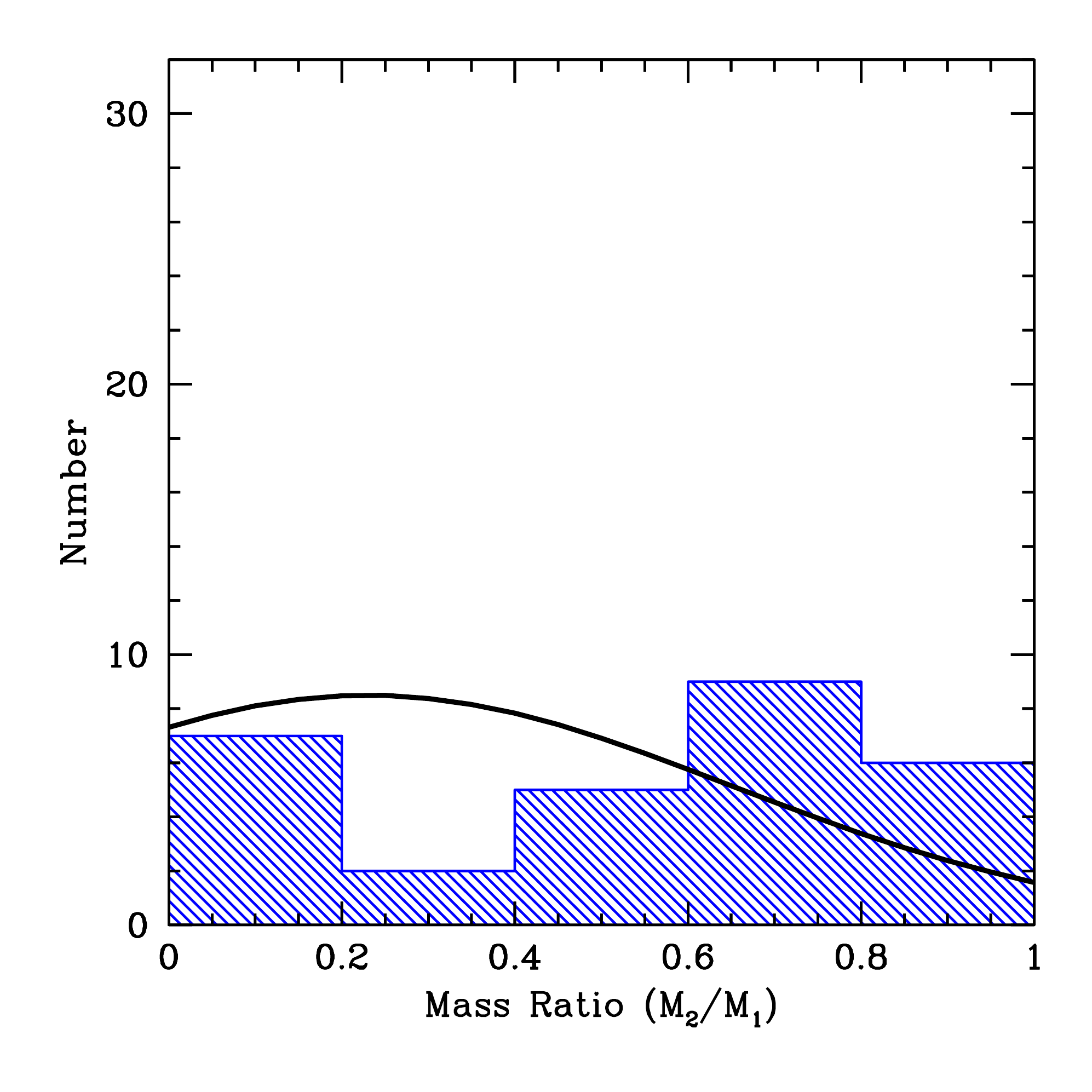}
    \includegraphics[width=5.0cm]{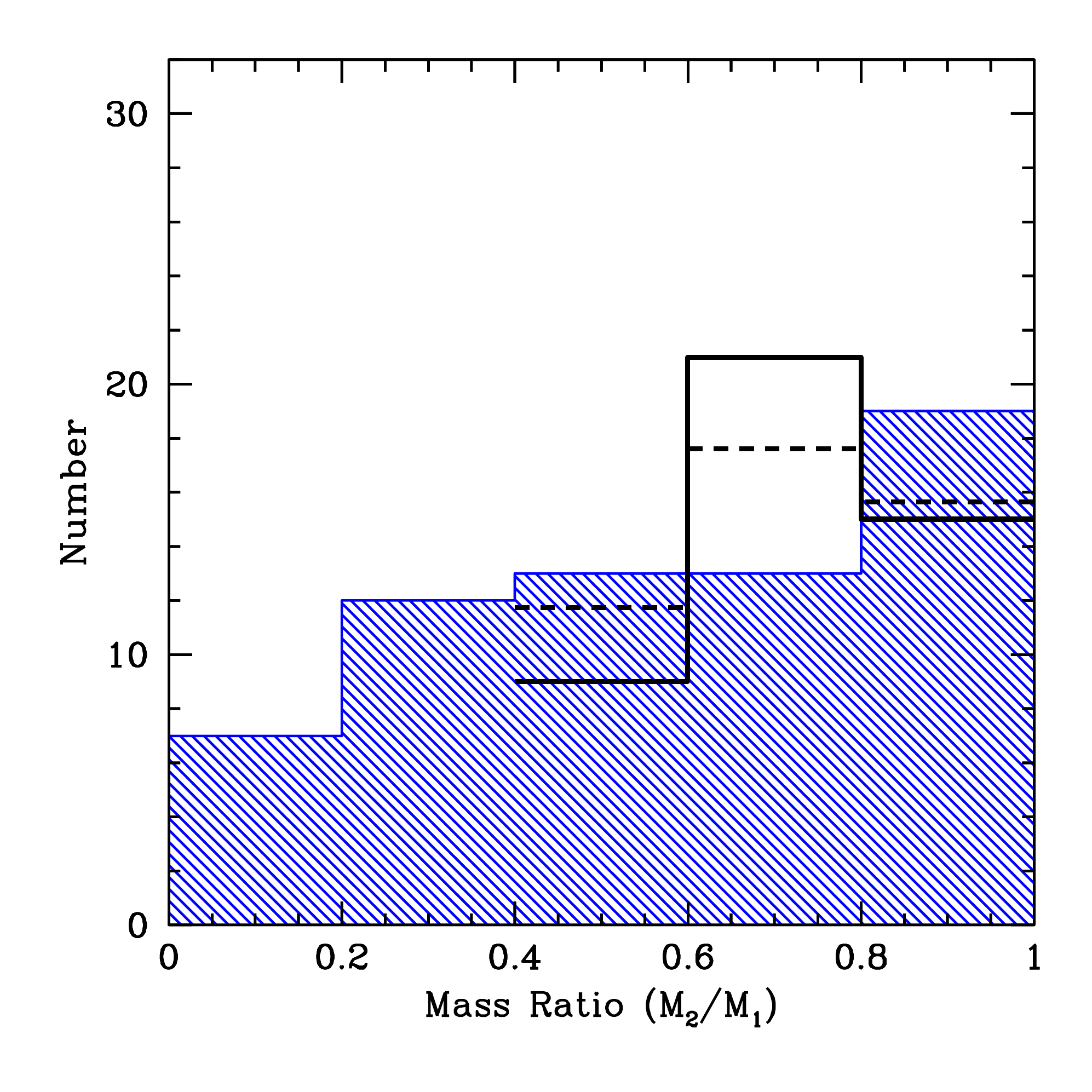}
    \includegraphics[width=5.0cm]{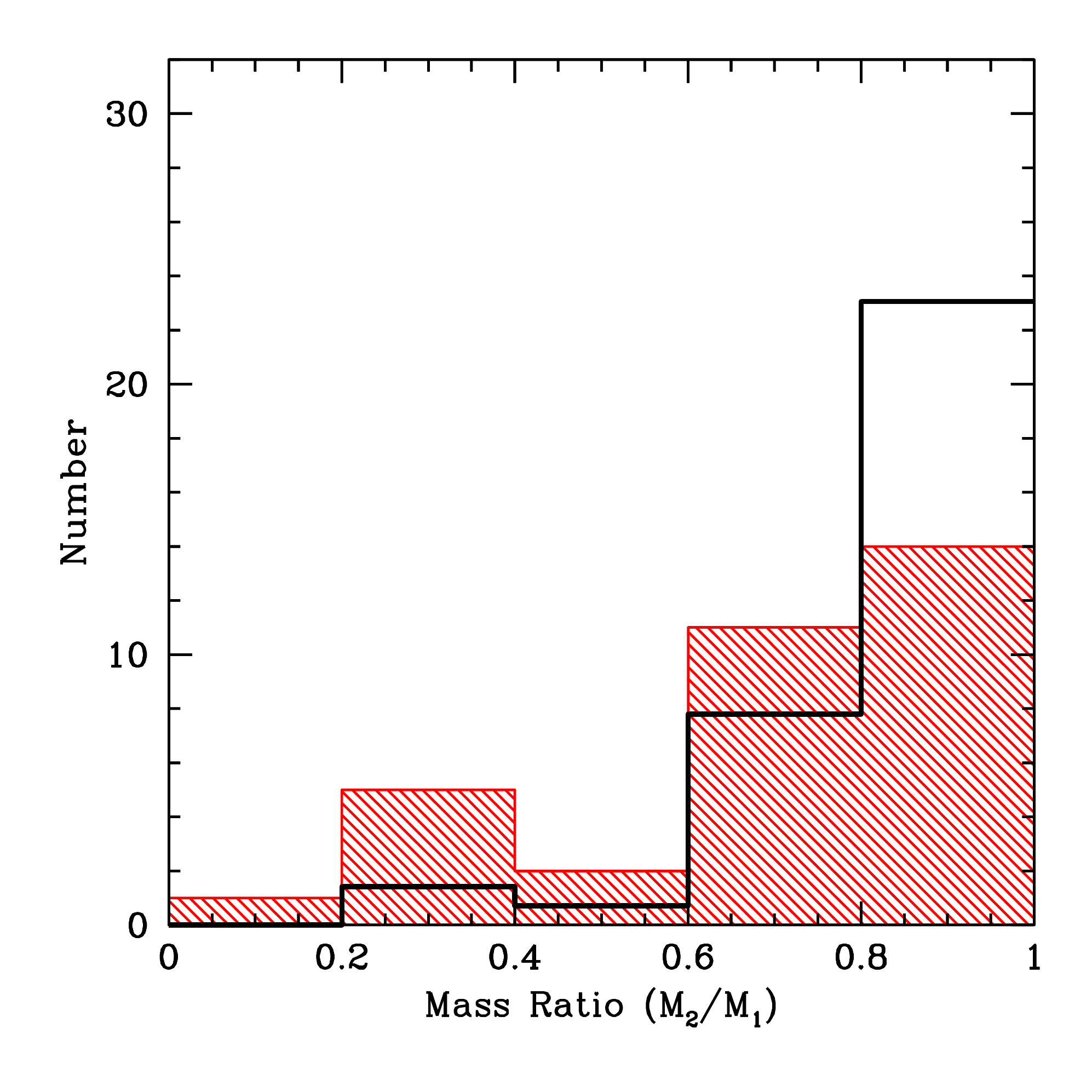}\vspace{-5pt}
\caption{The mass ratio distributions of binary systems with stellar primaries in the mass ranges $M_1>0.5$ M$_\odot$ (left) and $M_1=0.1-0.5$ M$_\odot$ (centre) and VLM primaries (right; $M_1<0.1$ M$_\odot$).  The top panels give the results at the end of the hydrodynamical calculation (0.3~Myr; Bate 2009), while the results at the end of the {\it N}-body evolutions (10~Myr) with instantaneous gas removal and no gas removal are given in the middle and lower panels, respectively.  The solid black lines give the observed mass ratio distributions of \citet{DuqMay1991} for G dwarfs (left), \citet{FisMar1992} for $M_1=0.3-0.57$ M$_\odot$ (centre, solid line) and $M_1=0.2-0.57$ M$_\odot$ (centre, dashed line), and of the known VLM binary systems maintained by N. Siegler, C.~Gelino, and A.~Burgasser at http://vlmbinaries.org/ (right).  The observed mass ratio distributions have been scaled so that the areas under the distributions ($M_2/M_1=0.4-1.0$ only for the centre panels) match those from the simulations.  In all cases, VLM binaries are biased towards equal masses when compared with M dwarf binaries (primary masses in the range $M_1=0.1-0.5$ M$_\odot$).  At 10~Myr with instantaneous gas removal, 75\% of the VLM binaries have $M_2/M_1>0.6$ while for the M dwarf binaries the fraction is only 55\%. There is only weak evolution of the VLM and M-dwarf mass ratio distributions during the {\it N}-body evolution, even in the most extreme case of no gas removal.  However, with fast gas removal the {\it N}-body evolution results in a significant increase in the fraction of unequal mass solar-type binaries (the fraction of systems with mass ratios $M_2/M_1<0.4$ increases from 26\% to 42\% in the case with instantaneous gas removal).}
\label{massratios}
\end{figure*}

In the lower two rows of Figure \ref{separations}, we present results from the {\it N}-body calculations with gas potentials at 10~Myr.  The third row presents the results from the calculation with a gas removal timescale of $T_{0.1}=1$~Myr, while the bottom row presents the results from the calculation with no gas removal.  The $T_{0.1}=0.3$~Myr case has been omitted because, as might be expected, the results are nicely bracketed by the cases of instantaneous gas removal and $T_{0.1}=1$~Myr.  As expected, longer gas removal timescales destroy more wide systems ($100-10^4$~AU), with triple and quadruple systems preferentially disrupted.  However, there is very little difference between the calculations with different gas removal timescales at separations less than 100 AU and the formation of significant numbers of very wide systems (separations $>10^4$~AU) discussed above still occurs even without gas removal.  For the VLM systems, there is also very little dependence on the gas removal timescale.  In all cases, the main evolution is that the number of systems with separations of $100-10^4$~AU decreases substantially and, in all but the no gas removal case, a few very wide ($>10^4$~AU) systems are formed in the halo.  Even in the extreme case of no gas removal at 10~Myr, the separation distribution is in good agreement with observations.

\subsubsection{Mass ratio distributions}

As we have seen, subsequent dynamical evolution of the \cite{Bate2009a} cluster results in the formation of very wide systems and evolution of the separation of VLM binaries that is in good agreement with that observed.  How do the mass ratio distributions evolve?

In Figure \ref{massratios}, we present the mass ratio distributions of binaries with primaries that have masses $\geq 0.5$ M$_\odot$ (left panels), are M-dwarfs with masses $0.1\leq M<0.5$ M$_\odot$ (centre panels), or are VLM objects (right panels).  Again, the top panels are at the end of the hydrodynamical calculation (see also Figure 19 in \cite{Bate2009a}), while the other panels give the results at the end of the {\it N}-body evolutions at an age of 10 Myr.  The case with instantaneous gas removal is given by the middle panels, while the lower panels give the results with no gas removal.  We compare the M-dwarf mass ratio distribution to that of \citet{FisMar1992}, and the higher mass stars to the mass ratio distribution of solar-type stars obtained by \citet{DuqMay1991}.  The VLM mass ratio distribution is compared with the listing of VLM multiples maintained by N.\ Siegler, C.~Gelino, and A.~Burgasser at http://vlmbinaries.org/ (last updated July 28, 2009).  The figure only includes the mass ratios of binaries, but we include binaries that are components of triple and quadruple systems.  A triple system composed of a binary with a wider companion contributes the mass ratio from the binary, as does a quadruple composed of a triple with a wider companion.  A quadruple composed of two binaries orbiting each other contributes two mass ratios - one from each of the binaries.

As noted by \citeauthor{Bate2009a}, at the end of the hydrodynamical calculation, the ratio of near-equal mass systems to systems with dissimilar masses decreases going from VLM objects to M dwarfs in a similar way to the observed mass ratio distributions, but that the trend is not as strong as in the observed systems.  Specifically, 71\% of the VLM binaries had $M_2/M_1>0.6$ while for primary masses $0.1-0.5$ M$_\odot$ the fraction was only 51\%.  The M-dwarf mass ratio distribution is consistent with \citeauthor{FisMar1992}'s distribution.  The VLM binaries, although biased towards equal-mass systems, are not as strongly biased as is observed.  
The main problem with the hydrodynamical binary mass ratio distributions, however, is that amongst `solar-type' binaries with primary masses greater than 0.5 M$_\odot$ the proportion with mass ratios $M_2/M_1<0.5$ is much lower than that found by \citeauthor{DuqMay1991}.

At the end of the {\it N}-body evolution with instantaneous gas removal, the mass ratio distributions of the VLM and M-dwarf binaries have not changed significantly.  75\% of the VLM binaries have $M_2/M_1>0.6$ and 55\% of the M-dwarfs.  However, there is a substantial change in the mass ratio distribution of solar-type binaries.  The number of near-equal mass binaries has decreased somewhat, while the number of unequal mass binaries has substantially increased.  Whereas only $9/34=26$\% had $M_2/M_1<0.4$ at the end of the hydrodynamical calculation, the fraction is $15/36 = 42$\% at the end of the {\it N}-body evolution.

As with the evolution of the separation distributions, the evolution of the mass ratio distributions discussed above is qualitatively similar for all the randomly-perturbed realisations.  The number of unequal-mass solar-type binaries displays some variation.  In the unperturbed case there were 15 with $M_2/M_1<0.4$ (out of 36), while the mean value is slightly lower at $10.7 \pm 2.3$, and the minimum and maximum values are 7 and 15, respectively.  There is also some variation in the number of unequal-mass VLM binaries.  The unperturbed case left 9 with $M_2/M_1<0.6$ (out of 36), while the mean value is $8.7 \pm 2.0$ and the minimum and maximum values are 6 and 12, respectively.

The mass ratio distributions resulting from the {\it N}-body calculations with gas potentials also differ very little from those obtained with instantaneous gas removal.  Even the extreme case of no gas removal differs only slightly from the instantaneous case (compare the middle and lower rows of Figure \ref{massratios}).

To illustrate what has caused the increase in unequal-mass solar-type binaries in the case with instantaneous gas removal, we discuss the formation of the three binaries with mass ratios $0.2 < M_2/M_1 < 0.4$.  One of these systems was formed when a triple consisting of a 2.5 AU binary containing 0.96 and 0.73 M$_\odot$ stars with an 0.21~M$_\odot$ companion at 12 AU broke up to leaving the 0.73 and the 0.21 M$_\odot$ stars in a tight (0.1 AU separation) binary with a low-mass ratio.  Another was formed when a quadruple system consisting of two binaries with 21~AU and 5~AU semi-majors axes orbiting each other with a 95-AU semi-major axis broke up forming a new binary with a mass ratio of $M_2/M_1=0.31$ consisting of one component from each of the original binaries in a 12-AU semi-major axis.  The last was formed when 1.74 and 0.40~M$_\odot$ stars that had been ejected from the cluster with similar velocities happened to find themselves weakly mutually bound in a 4100-AU binary.  As was seen in the previous section, a significant number of wide systems are formed in this manner and they tend to contain (although not exclusively) of one of the more massive stars ($>0.5$~M$_\odot$), as might be expected for a wide pair of objects to be bound.  Another low-mass ratio example consists of 0.70 and 0.10~M$_\odot$ stars in a 7500-AU binary.

\begin{figure}
\centering
    \includegraphics[width=5.0cm]{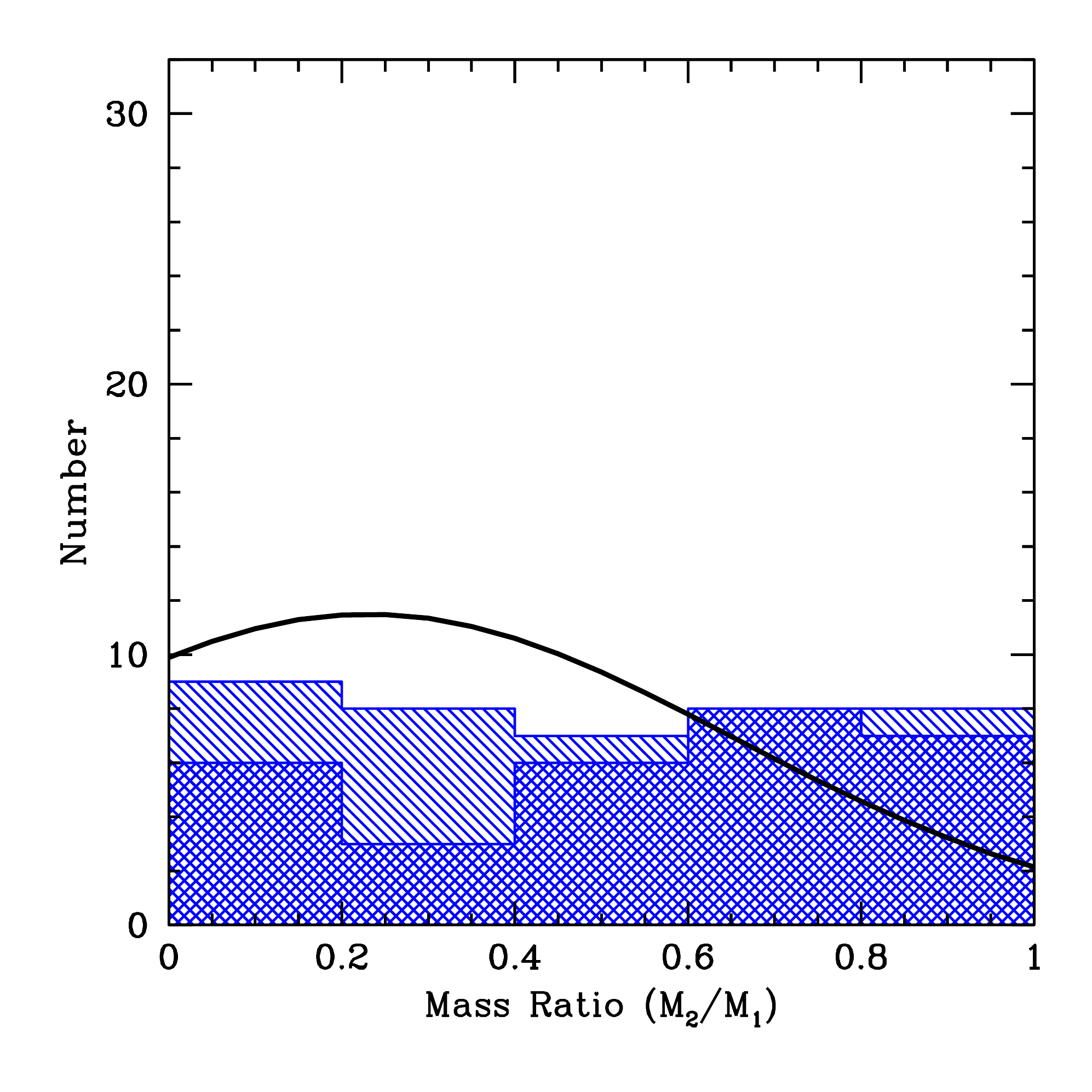}
\caption{The mass ratio distributions of binary, triple and quadruple systems with stellar primaries with masses $M_1>0.5$ M$_\odot$ at the end (10~Myr) of the unperturbed {\it N}-body evolution with instantaneous gas removal, excluding VLM companions (masses $<0.1$~M$_\odot$.  The double hashed region gives the mass ratio distribution for binaries only.  The single hashed region includes the mass ratios of triple and quadruple systems taking the mass ratio of the two most widely separated components.  The solid black line gives the fit to the observed mass ratio distribution of Duquennoy \& Mayor (1991) for G dwarfs.  The observed mass ratio distribution has been scaled so that the areas under the distribution matches that from the simulation.  }
\label{massratio_solar}
\end{figure}

The formation of low-mass ratio solar-type binaries during the {\it N}-body evolution with instantaneous gas removal goes some way to addressing the problem that hydrodynamical simulations of star formation appear to under-produce unequal-mass solar-type binaries \citep{Delgadoetal2004,Bate2009a,Clarke2009}.  \cite{Bate2009a} suggested that if the mass ratios of triples and quadruples were included this might also help to address the disagreement (i.e. essentially arguing that some of \citeauthor{DuqMay1991}'s unequal mass wide binaries might have in fact been unidentified triples or quadruples).  In Figure \ref{massratio_solar}, we plot the mass ratio distribution at the end of the {\it N}-body evolution of systems with primary masses $>0.5$~M$_\odot$ including the mass ratios between the wide components of triples and quadruple systems, but excluding VLM companions.  For example, for a triple, the mass ratio is that between the close binary and the wider companion.  For a quadruple consisting of two binaries, we include the mass ratio of the two binaries.  We exclude VLM companions because the survey of \cite{DuqMay1991} was not sensitive to VLM objects.  The inclusion of triple and quadruple systems, in addition to the formation of unequal-mass binaries during the {\it N}-body evolution, does not completely remove the discrepancy between the simulation and the survey of \cite{DuqMay1991}, but it does go a long way towards addressing it.  We also note that the recent surveys of binaries with separations $5-5000$~AU in star-forming regions \citep{Krausetal2008,KraHil2009} find uniform mass ratio distributions for primary masses $\gsim 0.75$~M$_\odot$, unlike that of \cite{DuqMay1991}.  However, \cite{KraHil2009} also found an indication of a variation between different star-forming regions with 9/12 of the wide binaries ($500-5000$ AU) in Taurus having $M_2/M_1>0.75$ while in Upper Sco 6/11 of the binaries had $M_2/M_1<0.25$.  In summary, the mass ratio distribution of binaries warrants further investigation, both observationally and theoretically.

\section{Discussion}

For the first time, we have studied the evolution following gas dispersal of a stellar cluster produced from a hydrodynamical calculation of the collapse and fragmentation of a turbulent molecular cloud.  Since only one hydrodynamical star formation calculation that forms a substantial cluster ($\gg 100$ stars) and resolves binary and multiple stellar systems has ever been performed \citep{Bate2009a} our results are, by necessity, limited to one particular type of cluster in terms of total mass, stellar density, etc.  However, while it is difficult to extrapolate our results more widely (e.g. to more or less massive clusters), the calculations presented here at least illustrate the types of evolution that are likely to occur in other types of clusters.

\subsection{Cluster evolution during gas dispersal}

The cluster studied here results from the hydrodynamical collapse and fragmentation of a dense turbulent molecular cloud and, just before gas dispersal, it is very compact (half-mass radius 0.05~pc), with a very high central stellar density ($\sim 10^6$~pc${^3}$), surrounded by a halo of ejected objects.  Despite this initial state, upon removal of the gas (which makes up 62\% of the total mass), a bound remnant cluster, that contains about 30-40\% of the mass and number of stars, expands to a radius of $\approx 1-2$~pc over a period of $\approx 4-10$~Myr (depending on the gas removal timescale).  This evolution is consistent with the many past {\it N}-body studies of the effects gas explusion on cluster evolution which have started from idealised initial conditions.  For example,  \cite{BauKro2007} performed a wide variety of {\it N}-body simulations of the cluster dissolution following gas dispersal with varing star formation efficiencies.  They found that their surviving clusters expanded by up to a factor of 10 in size.  The obvious question raised by this sort of evolution is whether or not real stellar clusters may originate from such compact configurations?

The amount of gas removed from the cluster we study here is large and only $30-40$\% of the total number of stars remain in the remnant cluster.  However, these numbers are not unreasonable. A star formation efficiency of $\approx 40$\% has been inferred for the ONC \citep{HilHar1998}, while \cite{Weidneretal2007} find observationally that low-mass clusters keep at most 50\% of their stars during gas loss. 

Recently, the observational evidence for significant expansion of young clusters has also been mounting.  \cite{ScaClaMcC2005} showed that if the current estimates of the virial ratio of the ONC (which show that the cluster is unbound) are correct, the size and age of the ONC imply either that it became unbound only very recently, or else that it has expanded quasi-statically. In the latter case, they infer that its initial central density may have exceeded its current value of $\sim 10^4$~pc$^{-3}$ \citep{McCSta1994,HilHar1998} by 1-2 orders of magnitude which would make it similar to the cluster studied here.  \cite{Parkeretal2009} also argue that the central region of the ONC was 2 orders of magnitude denser in the past with a half-mass radius of only 0.1-0.2~pc based on its current population of wide binaries.  Moving to clusters other than the ONC, \cite{Bastianetal2008} studied several dozen Galactic and extragalactic stellar clusters and found evidence for rapid expansion of the cluster cores during the first 20~Myr of their evolution.  Examining the evolution of cluster sizes with age, they found that clusters may begin with core radii $\lsim 0.1$~pc in size. 

How else might such young cluster evolution be tested?  From their {\it N}-body simulations, \cite{BauKro2007} found that if gas explusion is rapid, the stars acquire strongly radially anisotropic velocity dispersions outside their half-mass radii.  Similarly, in the cluster evolution presented here, the ejected halo population which comprises $>60$\% of the stars and stellar mass are on unbound essentially radial trajectories with velocities that increase with distance from the cluster remnant (faster moving stars have travelled further).

Finally, as noted in Section \ref{LagrangianSection}, a common feature of the {\it N}-body calculations is for the most massive binary to be ejected from the remnant cluster, sometimes accompanied by a small group of $\approx 20$ companion stars.  Many Herbig Ae/Be stars are associated with small groups of stars \citep{Hillenbrand1994,Hillenbrand1995,Hillenbrandetal1995}.  Since the ejection of such a group occurs in the several of our randomly-perturbed realizations, we suggest that this might be a significant formation mechanism of Herbig Ae/Be stars in small groups as opposed to an intermediate-mass star forming with a few other stars in an isolated molecular cloud.


\subsection{The evolution of binary and multiple stellar systems}

As noted in Section \ref{introduction}, many previous {\it N}-body simulations have investigated the evolution of binary properties, particularly binary frequency and separation distributions, during the early evolution of a young cluster \citep[e.g][]{Kroupa1995a,Kroupa1995b,Kroupa1998,Parkeretal2009}.  Many of the conclusions reached from these studies are also applicable to the cluster we have studied here.  In particular, we find a cluster remnant surrounded by a halo of ejected stars with a lower binary fraction \citep{Kroupa1995b}.  We also find that some of the most massive stars can be ejected and reach large distances from the cluster within 10~Myr \citep{Kroupa1998}.  However, unlike \cite{Kroupa1995a} and \cite{Parkeretal2009} we do not find that many primordial binary systems are disrupted.  Rather, the vast majority of binary systems formed during the star formation process (i.e. the hydrodynamical evolution) survive the gas dispersal phase with little evolution of binary frequency or their separation distribution.  This is because stars do not form in the hydrodynamical simulations with 100\% of them being in binaries, nor do they form with a flat distribution in log-separation.  Rather, star formation and dynamical interaction cannot be separated -- they proceed simultaneously, resulting in a `primordial' binary population that is already somewhat stable against further evolution.  To try to separate star formation and the properties of primordial binary properties is likely to severely over-estimate the importance of the dynamical evolution of the binary population early in the life of a stellar cluster.

By contrast, we note that higher-order multiple systems (which are usually not included in purely {\it N}-body studies at all) do undergo significant evolution during the early evolution of a stellar cluster.  In particular, many quadruple systems that result from the star formation process are found to decay into binaries or triples.  Thus, we assert that observational studies of higher-order multiple systems in very young stellar clusters are likely to find a much higher proportion of high-order multiples than is found in open clusters or the field population \citep[e.g.][]{Reipurth2000}.

Finally, as presented in Section \ref{multiplicitysection}, we find that a significant fraction of very wide (separations $10^4-10^5$~AU) multiple systems (around half of which are actually triples or quadruples) are produced in the halo of objects ejected from the cluster.  This has also been noted in recent pure {\it N}-body simulations of dispersing clusters \citep{Kouwenhovenetal2009}. Thus, even though the hydrodynamical star formation process in a very dense cluster such as that discussed here (half-mass radius only 0.05~pc, or $10^4$~AU at the end of the star formation)  cannot not produce very wide systems by direct fragmentation, such wide systems can be produced during cluster dissolution in numbers that are large enough to explain the observed field population of wide systems \citep{DuqMay1991}.  Such formation of wide systems during cluster dissolution potentially solves the problem \citep{Parkeretal2009} of how a broad log-separation distribution \citep[e.g.][]{DuqMay1991} can be obtained if most stars form in clusters \citep{AdaMye2001,LadLad2003} which cannot contain wide binaries. Moreover, many of these wide binary systems and the binaries formed from the decay of higher-order multiple systems have unequal mass components.  This also helps to fix the deficit of unequal-mass solar-type binaries (relative to the observations of \citealt{DuqMay1991}) that is usually present at the end of the hydrodynamical star formation phase \citep{Delgadoetal2004,Bate2009a,Clarke2009}.  Thus, in summary, the formation of significant numbers of wide multiple systems during cluster dissolution potentially solves several problems simultaneously.

\section{Conclusions}

We have presented the results of {\it N}-body calculations that take the end point of the hydrodynamical star cluster formation calculation of \cite{Bate2009a} and evolve the stellar cluster to an age of 10~Myr. The calculations are unique in the sense that they begin from a star cluster that has self-consistently formed from the hydrodynamical collapse and fragmentation of a molecular cloud to form a stellar cluster containing binary and higher-order multiple systems rather than beginning from arbitrary initial conditions.  Although we are limited to studying one particular type of cluster in terms of the number of stars, stellar density, etc, we find many of the same evolutionary processes that have been studied in past {\it N}-body calculations that begin from arbitrary initial conditions.  For example, we find that a bound remnant cluster containing $\approx 30-40$\% of the stars by number or mass remains despite the low star formation efficiency (38\%) and this cluster is surrounded by a halo of ejected stars.  The remnant cluster expands from an initial radius of less than 0.05~pc to $\approx 1-2$~pc over a period of $\approx 4-10$~Myr (depending on the gas removal timescale).  This result, along with other recent numerical and observational studies, suggest that young clusters may begin from much higher stellar densities ($10^5-10^7$~pc$^{-3}$) than usually assumed.

When investigating mass segregation in our young clusters, we differentiate between primordial mass segregation, which is a result of the star formation process, and classical mass segregation, which is due to dynamical relaxation.  We find that primordial mass segregation is only evident for the few most massive stars (in agreement with observations of the Orion Nebula Cluster), but only lasts a short time ($\approx 1$~Myr).  Later, classical mass segregation may develop with higher-mass stars being increasingly centrally condensed, but this depends on the gas removal timescale.  Slower gas removal results in shorter relaxation times and, thus, more classical mass segregation.

Although many of the global cluster evolutionary phases have been discussed before in the literature, we also find some new evolutionary processes at work.  We find that in the majority of our calculations, some of the most massive stars in the cluster are ejected from the bound cluster and are accompanied by a small group of companion stars.  Thus, we propose that this mechanism might be responsible for the formation of Herbig Ae/Be stars that are accompanied by small stellar groups.

Turning to the evolution of binary and multiple systems, we note that past {\it N}-body studies of the evolution of young clusters that begin with high binary frequencies and wide separation distributions have emphasised the rapid destruction of binary systems.  However, in our calculations the binary frequency remains almost unchanged throughout the {\it N}-body evolution.  We emphasise that the processes of star formation and the formation of multiple stellar systems cannot be separated during the formation of a cluster.  Instead, they occur simultaneously, resulting in a `primordial' binary population that is already somewhat stable against further evolution.  That said, we do find that the higher-order multiple systems produced by the star formation process, particularly quadruples, evolve significantly with most decaying into lower-order systems.  This is potentially observable by comparing the frequencies of triple and quadruple systems in young star-forming regions with those found in open clusters.  We also find very wide ($10^4-10^5$~AU) binary and higher-order multiple systems are formed in the dispersing halo of ejected objects when objects happen to be expelled with similar velocities and find themselves weakly bound.  These wide binaries are formed in sufficient numbers to explain the observed separation distribution of solar-type stars despite the fact that the original star-forming cloud is too small to form such systems by direct fragmentation.  Finally, whereas the cluster at the end of the hydrodynamical evolution was deficient in unequal-mass solar-type binaries compared with observations, we find that many of the wide binaries formed in the halo and those binaries resulting from the decay of higher-order multiple systems have unequal mass components.  Thus, these processes may help to bring the mass ratio distributions of binaries produced by hydrodynamical simulations into agreement with the observed mass ratio distributions of solar-type stars.

\section*{Acknowledgments}

MRB is grateful for the support of a EURYI Award and for the hospitality and financial support provided by the Isaac Newton Institute for Mathematical Sciences as part of its Dynamics of Discs and Planets programme where part of this paper was written up.  This paper has made use of the Very Low Mass Binaries archive (http://vlmbinaries.org/) maintained by N.~Siegler, C.~Gelino, and A.~Burgasser.  This work, conducted as part of the award ÒThe formation of stars and planets: Radiation hydrodynamical and magnetohydrodynamical simulationsÓ made under the European Heads of Research Councils and European Science Foundation EURYI (European Young Investigator) Awards scheme, was supported by funds from the Participating Organisations of EURYI and the EC Sixth Framework Programme.

\bibliography{mbate}

\end{document}